\documentclass[sigconf]{acmart}

\usepackage{fancyhdr}

\acmConference[ASE'24]{ACM Conference}{Oct 2024}{Sacramento, USA}

\usepackage{tabularx}
\usepackage{marvosym}
\usepackage{multicol}
\usepackage{multirow}
\usepackage{graphicx}
\usepackage[framemethod=TikZ]{mdframed}

\usepackage{colortbl}
\usepackage{subcaption}
\usepackage{array}
\usepackage{float}
\usepackage{caption}
\usepackage{wrapfig}

\copyrightyear{2024}
\acmYear{2024}
\setcopyright{acmlicensed}\acmConference[ASE '24]{39th IEEE/ACM International Conference on Automated Software Engineering }{October 27-November 1, 2024}{Sacramento, CA, USA}
\acmBooktitle{39th IEEE/ACM International Conference on Automated Software Engineering (ASE '24), October 27-November 1, 2024, Sacramento, CA, USA} \acmDOI{10.1145/3691620.3694987} \acmISBN{979-8-4007-1248-7/24/10}

\newcolumntype{C}[1]{>{\centering\arraybackslash}p{#1}}

\AtBeginDocument{%
  }

\begin{document}

\title{Preference-Guided Refactored Tuning for Retrieval Augmented Code Generation}

\author{Xinyu Gao }
\affiliation{%
  \institution{Shanghai Key Laboratory of Data Science, School of Computer Science,}
  \institution{Fudan University}
  \city{Shanghai}
  \country{China}}
\email{23210240161@m.fudan.edu.cn}

\author{Yun Xiong \textsuperscript{\Letter}}

\affiliation{%
  \institution{Shanghai Key Laboratory of Data Science, School of Computer Science,}
  \institution{Fudan University}
  \city{Shanghai}
  \country{China}}
\email{yunx@fudan.edu.cn}

\author{Deze Wang}
\affiliation{%
  \institution{National University of Defense Technology}
  \city{Hunan}
  \country{China}}
\email{wangdeze14@nudt.edu.cn}

\author{Zhenhan Guan}
\affiliation{%
  \institution{Shanghai Key Laboratory of Data Science, School of Computer Science,}
  \institution{Fudan University}
  \city{Shanghai}
  \country{China}}
\email{zhenhan_guan@163.com}

\author{Zejian Shi}
\affiliation{%
  \institution{Shanghai Key Laboratory of Data Science, School of Computer Science,}
  \institution{Fudan University}
  \city{Shanghai}
  \country{China}}
\email{zjshi20@fudan.edu.cn}

\author{Haofen Wang}
\affiliation{%
  \institution{College of Design and Innovation,}
  \institution{Tongji University}
  \city{Shanghai}
  \country{China}}
\email{carter.whfcarter@gmail.com}

\author{Shanshan Li \textsuperscript{\Letter}}
\affiliation{%
  \institution{National University of Defense Technology}
  \city{Hunan}
  \country{China}}
\email{shanshanli@nudt.edu.cn}

\begin{abstract}

Retrieval-augmented code generation utilizes Large Language Models as the generator and significantly expands their code generation capabilities by providing relevant code, documentation, and more via the retriever. The current approach suffers from two primary limitations: 1) \textbf{information redundancy.} The indiscriminate inclusion of redundant information can result in resource wastage and may misguide generators, affecting their effectiveness and efficiency. 2) \textbf{preference gap.} Due to different optimization objectives, the retriever strives to procure code with higher ground truth similarity, yet this effort does not substantially benefit the generator. The retriever and the generator may prefer different golden code, and this gap in preference results in a suboptimal design. Additionally, differences in parameterization knowledge acquired during pre-training result in varying preferences among different generators.

To address these limitations, in this paper, we propose \textbf{RRG} (\underline{\textbf{R}}etrieve, \underline{\textbf{R}}efactor, \underline{\textbf{G}}enerate), a novel framework for effective and efficient code generation. This framework introduces a code refactorer module between the retriever and the generator to bridge them. The refactoring process transforms the raw retrieved code into a more concise, efficient, and model-friendly version. It eliminates redundant information and noise, reducing the input length. Consequently, the generator receives higher-quality context, enabling it to produce more accurate results with lower inference costs. We conducted comprehensive experiments on multiple datasets. In the experiments, we confirmed the existence of a preference gap between the retriever and the generator, and RRG effectively bridges this gap. Specifically, RRG achieved significant performance improvements, with increases of up to 28\% on EM, 13\% on BLEU, and 6.8\% on CodeBLEU.

\end{abstract}
\maketitle

\begin{figure}[]
    \centering
    \includegraphics[scale = 0.65]{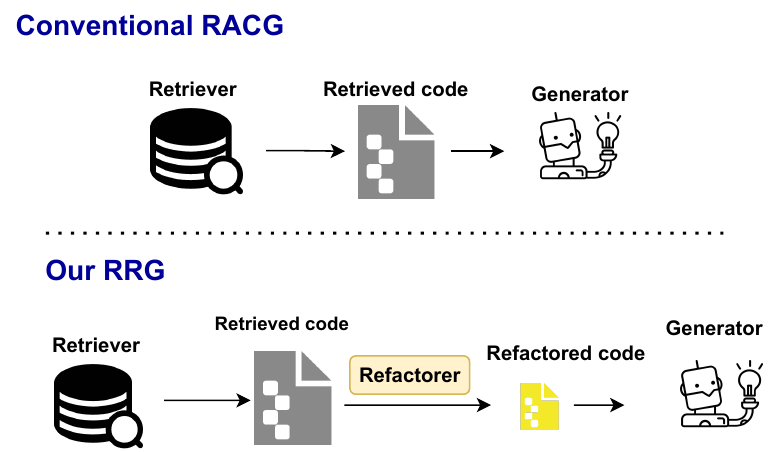}
    \caption{Comparison of two RACG frameworks. The upper shows the conventional approach, where the retrieved code is directly fed to the LLMs directly, but this holds redundant information and is not model-friendly. The lower shows the RRG framework, where the retrieved code is refactored into concise, model-friendly code, which improves the effectiveness and efficiency of the generator. }
    \label{intro-1}
\end{figure}

\section{Introduction}
Large Language Models (LLMs) \cite{reid2024gemini,codegeex,shen2023pangu,tian2023chatgpt} have attracted significant attention due to their extraordinary capabilities. Many studies \cite{generaion1,generation2,generaion3} have applied them to code generation tasks with impressive performance. However, the rapid versioning of code repositories and the large syntactic gaps between programming languages make it difficult to meet expectations for LLMs solely based on in-parametric knowledge \cite{codegen4lib,ase-gold}. Retrieval-augmented code generation (RACG) \cite{recoder,docprompting,arks} significantly enhances the code generation capability of LLMs by utilizing them as generators and providing relevant knowledge through the retriever. Early explorations have alleviated the generator's knowledge updating and long-tail \cite{long-tail,zhao2024ltgc} problems to some extent by searching code repositories or retrieving solutions via Internet search engines \cite{recoder,docprompting}.

However, the existing approach suffers from the following limitations:

 \textbf{\romannumeral1) Information redundancy.} Due to the inherent limitations of the retriever or external databases, some superfluous information is inevitably retrieved. If such redundant information is directly incorporated into the generator's context, it may mislead the generator \cite{Less-is-More,inrelevant}, and potentially introduce misleading or harmful information, thereby impairing the generator's overall performance. Additionally, the retriever may retrieve an excessive volume of related documents or code. This not only increases inference costs and risks getting lost in middle \cite{lost}, but also results in the loss of critical information due to exceeding the context length limit for smaller models. Previous researches have explored removal of redundant information from the perspectives of information entropy and relevance \cite{xu2023recomp,token_elimination,crag,jiang-etal-2023-llmlingua}. Generally, tokens with lower perplexity are deemed less crucial for the model's contextual understanding \cite{li2023unlocking}, and tokens with lower relevance and perplexity are filtered out. This does not consider the relationship between code tokens, and the compressed code depends entirely on the raw code \cite{pan2024llmlingua}. In the RACG pipeline, if the retrieved non-parameterized knowledge lacks relevant code, selective compression also fails to provide the generator with relevant code.

\textbf{\romannumeral2) Preference gap.} From a human perspective, documents containing correct answers are the most helpful in RACG \cite{rag}. Therefore, the retriever is designed to retrieve the most relevant code to better assist the generator. However, our experiments reveal that code snippets more closely related to the reference answers do not necessarily enhance the generator's effectiveness, as shown by the results presented in Table \ref{gaps-exp}. The model may prefer a code snippet with a clear code structure, low repetition, and novel knowledge not already incorporated into its parameterized knowledge during the pre-training process \cite{devlin2018bert}. The difference between the golden code snippet preferred by the retriever and the generator is termed the preference gap. The existence of this preference gap renders the current RACG framework inadequate. Some studies \cite{shi2023replug,radit} have attempted to align them by jointly fine-tuning them or utilizing feedback signals from the generator. However, some LLMs accessed through APIs \cite{tian2023chatgpt} cannot be fine-tuned in conjunction with the retriever, nor do they provide logits to compute certain feedback signals. Additionally, changes in the semantic representation of text following the retriever parameter tuning necessitate updating the external database index, which consumes substantial resources in practice.

To address the aforementioned issues, we propose the \textbf{RRG} (\underline{\textbf{R}}etrieval, \underline{\textbf{R}}efactor, \underline{\textbf{G}}eneration) framework. We inserted a new module, a code refactorer, between the retriever and the generator. Fig. \ref{intro-1} shows the difference between our RRG and conventional RACG. This module processes the retrieved code, providing the generator with more efficient and model-friendly contextual information.

We propose a two-stage training scheme to train a preference-guided refactorer. In the first stage,we use Supervised Fine-tuning (SFT) to train the refactorer for generative compression, and during this process, we inject knowledge from an external knowledge database into the refactorer. Even if the retrieved information lacks validity, the refactorer can still rely on its internal parametric knowledge to generate information that is more relevant to the answer. In the second stage, we bridge the preference gap by tuning the parameters of the refactorer while freezing those of both the generator and the retriever. Specifically, we treat the generator as a reward model and the refactorer as a policy model, using reinforcement learning to optimize the refactorer’s output to better align with the generator's preference, guided by the generator's performance on downstream tasks. Notably, after the first stage of training, the refactorer can leverage both non-parametric and internal parametric knowledge to meet the generator's preference.

To comprehensively validate the performance of the RRG, we conducted extensive experiments across multiple datasets. The results demonstrate the remarkable flexibility and synergy of the RRG with any retrievers and generators, significantly enhancing the performance of the generator while reducing inference costs.
In summary, our contributions are as follows:
\begin{itemize}
    \item We propose a novel RACG framework, named RRG, achieved by incorporating a code refactorer module. Experimental results indicate that there indeed exists a gap between the retriever and the generator. This module compresses and optimizes the retrieved code, providing a more concise, efficient, and model-friendly context for the generator. Furthermore, it aligns this preference gap, thereby improving performance while reducing inference costs.
    \item  To train the preference-guided refactorer, we design a two-stage training scheme that first empowers the refactorer to perform generative compression using both parametric and non-parametric knowledge, and then optimizes the refactorer through feedback signals from the generator. This effectively bridges the preference gap between the retriever and the generator.
    \item Through extensive experiments on multiple datasets, we have verified that RRG significantly enhances the performance of generators of various types and scales in code generation tasks. This demonstrates its strong generalization capability and practical applicability.
\end{itemize}

\begin{figure*}[ht]
    \centering
    \includegraphics[scale = 0.5]{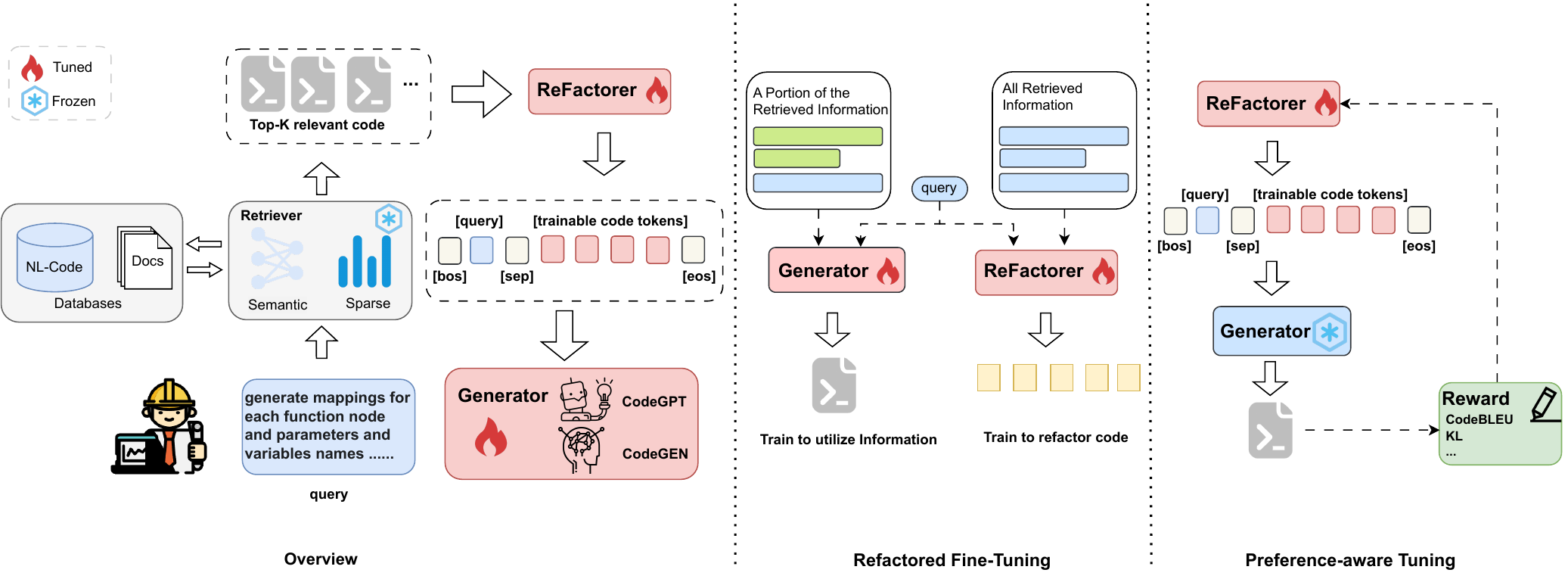}
    \caption{The overall architecture of our proposed RRG with Preference-Guided Refactored Tuning.}
    \label{overview}
\end{figure*}

\section{Related Work}

\subsection{Pretrained Models for Code Generation}

The evolution of LLMs has profoundly revolutionized the domain of code generation. While early LLMs \cite{gpt3,llama,t5} were not initially designed specifically for code generation, their pre-training incorporated a substantial amount of code corpus. This enabled them to exhibit remarkable proficiency in generating code. To further improve the performance of LLMs in code generation, various approaches \cite{codet5+,santacoder} have extended the pre-trained models using extensive code corpus. The code models like CodeLLaMA \cite{codellama} and CodeGeex \cite{codegeex} have also built upon their base models to significantly enhance performance in code generation and comprehension. CodeGEN \cite{codegen} takes a unique approach, perceiving code generation as a multi-round dialogue task, whereas AST-T5 \cite{gong2024ast} incorporates the syntactic intricacies of code into LLMs. These innovations underscore the vast potential of utilizing LLMs for code generation.

\subsection{Retrieval-Augmented Generation}

RAG \cite{radit,rag3} enhances LLMs by retrieving relevant knowledge from external knowledge databases, significantly improving the performance of LLMs in knowledge-intensive domains. Researchers continuously optimize the RAG pipeline from three perspectives: what, when, and how. \textbf{What} refers to what to use to enhance LLMs, such as Internet search engine \cite{webglm,webRAG}, knowledge graph \cite{rag-kg1,rag-kg2}, phrase \cite{rag-dy1,rag-dy2}, token \cite{rag-token1}, and chunk \cite{rag3}. \textbf{When} refers to when to retrieve. Due to the limitation of one-time retrieval at the beginning, Flare \cite{Active} proposes adaptive retrieval for more flexible and efficient knowledge acquisition. Self-RAG \cite{self-rag} enables the generator to reflect and improve its effectiveness through more comprehensive post-retrieval processing. \textbf{How} refers to how to use the retrieved information. RECOMP \cite{xu2023recomp} and PRCA \cite{prca} compress and extract the retrieved information, providing a more effective context for the generator. RETRO \cite{intro,wang2023shall} inputs the vectorized retrieved information into the internal of the model.

Additionally, BGM \cite{bgm} starts from the preferences of humans and LLMs, while we begin by studying the gap between the retriever and the generator. Additionally, BGM introduces a reranker, whereas our approach is more akin to a compressor module.

\subsection{Retrieval-Augmented Code Generation}

In the field of code generation, many studies on RACG have emerged in recent years. Classic works such as REDCODER \cite{recoder}, ReACC  , and DocPrompt [3] assist in tasks like code generation, summarization, and completion by retrieving external contexts and documents. Many studies further optimize naive RACG. KNN-TRANX \cite{knn-tranx} further optimized retrieval results through syntax awareness. LAIL \cite{lail} trained retrievers using signals from LLMs to align preferences. RepoFuse \cite{liang2024repofuse}  and APICoder \cite{code-api} reduced the retrieval granularity to private function libraries, enhancing the practicality of RAG. RepoCoder \cite{repocoder} optimized retrieval through iterative retrieval, though it incurred significant costs. Repoformer  and ARKS \cite{arks} improved efficiency and reduced costs through active retrieval. Toolcoder \cite{zhang2023toolcoder} and Codeagent \cite{zhang2024codeagent} apply the agent-based approach, allowing LLMs to use APIs and various tools to complete code generation.

Existing research in RACG has achieved notable progress. Nevertheless, it has overlooked addressing the preference gaps between the retriever and the generator, and it has failed to incorporate parameterized knowledge to guide the generator. In contrast, our work introduces a refactoring module designed to bridge the preference divide between the retriever and the generator. By harnessing both parameterized and non-parameterized knowledge, this module equips the generator with a richer, higher-quality contextual background.

\section{Approach}
\subsection{Overview}

We present our overall architecture and the two-phase training scheme in Fig. \ref{overview}. Given query $q$, the retriever will use a retrieval algorithm to fetch the top-K most relevant codes to $q$ from external knowledge databases. Subsequently, the code refactorer then converts the retrieved code into concise and model-friendly code. After concatenating the refactored code with $q$ using special tokens, the combined input is then fed to the generator to generate targeted code.

\textbf{External knowledge databases.} This can be either nl-code pairs or code documentation. We expect the external knowledge database to contain information related to the $q$.

\textbf{Retriever.} It retrieves the most relevant documents from external knowledge databases using a search algorithm, typically a vector retrieval algorithm. This generally requires two semantic representation models for a given corpus $C$ and $q$. The corpus encoder maps the natural language in $C$ to an n-dimensional vector, while the query encoder maps $q$ to an n-dimensional vector. Semantic similarity is measured based on the similarity between these vectors, which is typically calculated using the dot product.

\begin{equation}
    S(q,NL_{i}) = E_{corpus}(NL_{i})\cdot E_{query}(q)
\end{equation}

where $E_{corpus} $ and $E_{query}$ are the semantic representation models for corpus and queries respectively. And $NL_{i} $is the natural language
in (NL-code) pairs.

However, this approach faces the problem of unclear expressions and inaccurate semantic representations, necessitating the consideration of sparse algorithms based on word frequency. We use the BM25 algorithm to perform a second recall on the documents retrieved by semantic retrieval.

\textbf{Refactorer and Generator.} We introduce a code refactorer module between the retriever and the generator. This module processes the query $q$ along with the top-K relevant codes retrieved by the retriever, with the goal of extracting concise, model-friendly code snippets to better guide the generator. The generator, on the other hand, focuses on directly generating the target code based on the refined contextual information and query $q$. We aim for the refactorer to both act as a compressor and align the preferences of the retriever and the generator \cite{bgm}. To achieve this, we optimize the refactorer and the generator through a two-stage training scheme.

\subsection{Stage 1: Refactored Fine-Tuning }

During the refactored fine-tuning phase, we train the refactorer to generate compressed code based on human preference, and we train generators to use external knowledge for generating target code. In this section, we provide a detailed description of how to supervise the training of the refactorer and the generator.

\subsubsection{\textbf{Training to refactor code}}

The process of refactoring aims to condense the retrieved code to generate higher quality code snippets. At this stage, our focus is solely on refactoring the code, without considering alignment preferences. RACG is an open-book generative task. From a human perspective, providing relevant code snippets that closely resemble the target code significantly enhances the generator's performance. Consequently, we directly train the code refactorer to produce the target code, ensuring that the generated code closely aligns with our expectations. This condenses the retrieved code into a more concise form. Furthermore, we integrate knowledge from the query, relevant code, and target code into the refactorer, enabling it to produce code tokens pertinent to the query, even when the information retrieved by the retriever may not be perfectly relevant.

We formalize this task as a sequence-to-sequence code generation task, i.e., generative-compression. To realize the code refactorer, we employ encoder-decoder models such as T5 \cite{t5} and CodeT5 \cite{codet5,codet5+}. Such architectures perform well on summarization and generation tasks. 

Specifically, we concatenate the top-k retrieved codes with query $q$ via special tokens, where these k codes are also separated using special tokens and ranked based on their natural language relevancy to the query. Subsequently, the encoder model of the code refactorer takes in this sequence of tokens and generates an implicit vector representation that contains the contextual information of the original inputs, based on which the decoder generates tokens successively in an autoregressive mode. The actual output is a probability distribution of tokens, which is transformed into code tokens displayed in the vocabulary through a sampling strategy. The decoder continuously predicts the next token based on the generated tokens until the end-of-sequence token $[EOS]$ is generated. 

We use cross-entropy loss function to measure the disparity between the refactorer's output and the golden code, optimizing continuously to minimize this disparity, thus bringing the generated code closer to the expected target. The cross-entropy loss function is as follows:

\begin{equation}
Loss = -\frac{1}{T} \sum_{t=1}^{T} \sum_{i=1}^{V} y_t(i) \cdot \log(p_t(i))
\end{equation}

where \(T\) is the length of the sequence, \(V\) is the size of the vocabulary, and \(y_t(i)\) and \(p_t(i)\) are the probabilities of the \(i\)-th word in the vocabulary at time step \(t\) according to the true distribution and the model's predicted distribution, respectively.

\subsubsection{\textbf{Training to utilize information}}

To address context-based code generation tasks, we choose autoregressive models (e.g., GPT-3 \cite{gpt3}, CodeGPT \cite{codegpt}, etc.) as generators, which show excellent performance in generative tasks. In traditional code generation tasks, the generator produces the target code solely based on a given query. However, in the RACG framework, the input of the generator combines the query and the retrieved code.
In order to adapt to the specific input form, we fine-tuned the generator. In addition, in the RRG framework, to ensure that the generator matches the output of the refactoring, we limit the length of the relevant code in the SFT phase to the same maximum output length of the refactorings. By cropping the relevant code through a specifically set window, we can shorten the input context length for the model. This not only reduces the complexity of training but also allows the model to concentrate on extracting crucial information from the limited context, thus facilitating the generation of the target code.
Specifically, the input to the model contains the query $q$ and the cropped relevant code. First, we use the generator's tokenizer to process the relevant code with the splitter, and adjust its length to a fixed-length relevant-code$_{cropped}$ by cropping or padding operations. Subsequently, we concatenate the processed $q$, relevant-code-{cropped}, and the target code via a special token.

During training, we calculate the loss by comparing the probability distribution of the predicted next tokens at corresponding positions in the target code with the ground truth tokens. Specifically, we utilize the cross-entropy loss function to measure the discrepancy between the predicted and the ground truth  distributions.

\subsection{Stage 2: Preference-aware Tuning }

In the preference-aware-tuning phase, the refactorer is expected to align the preference of the retriever and the generator based on the non-parameterized knowledge and the internal parameterized knowledge. We treat the generator as a reward model and the refactorer as a strategy model \cite{ppo1}. We adjust the behavior of the strategy model based on the direct performance of the generator on downstream tasks. The code refactorer's action at a time step is to generate a token. However, using the reward model to give feedback on every token would be resource-intensive and could lose the global perspective, neglecting the relationship between tokens. Therefore, we take into consideration the start of the refactoring from the beginning of token generation to the generation of [EOS] token, i.e., the generation of a complete sentence, as the action of one time step of the refactorer \cite{ppo2}. To avoid training instability caused by the significant disparity between the initial refactorer's output and the generator's preference, we employ a PPO \cite{ppo} strategy that is more suitable for our scenario.

In Equation \ref{ppo_main}, the policy loss function of the PPO algorithm is presented. This is a clipped objective function that improves the training stability of the policy by limiting the changes made to the policy during each training epoch.

The ratio \( r_{t}(\theta) \) is defined as the probability of taking action \( a_{t} \) under the current policy \( \pi_{\theta} \) divided by the probability of taking the same action under the previous policy \( \pi_{\theta_{old}} \), expressed as: $ r_{t}(\theta ) =\frac{\pi_{\theta }(a_{t}|s_{t}) }{\pi_{\theta_{old}}(a_{t}|s_{t}) } $.
This ratio indicates the relative change in the likelihood of selecting an action between the current and previous policies. It is computed based on the probabilities \( \pi_{\theta}(a_{t}|s_{t}) \) and \( \pi_{\theta_{old}}(a_{t}|s_{t}) \), which represent the likelihood of taking action \( a_{t} \) given state \( s_{t} \) under the current and previous policies, respectively.

\begin{equation}
     L^{CLIP} (\theta )=\hat{\mathbb{E}}\left [ min(r_{t}(\theta )\hat{A_{t} },clip(r_{t}(\theta ),1-\varepsilon ,1+\varepsilon)\hat{A_{t} }) \right ]
     \label{ppo_main}
\end{equation}

where $r_{t}(\theta)$ is the policy ratio.The clipping operation restricts the policy ratio $r_{t}(\theta)$ within the range $[1-\varepsilon, 1+\varepsilon]$ to prevent large policy updates and ensure training stability.

$\hat{A}$ is an estimate of the advantage function. According to the original PPO \cite{ppo}, we generally use Generalized Advantage Estimation (GAE) to compute the advantage function. The GAE formula is as follows:

\begin{equation}
    \hat{A} ^{GAE(\gamma ,\lambda )} = \sum_{l=0}^{\infty } (\gamma \lambda )^{l}\delta _{t+l}^{V} 
\end{equation}

where $\gamma$ is the discount factor, typically used to measure the present value of future rewards. $\lambda$ is the smoothing factor in GAE, controlling the trade-off between bias and variance. $\delta_{t+l}^{V}$ is the Temporal Difference (TD) error at time step $t+l$, defined as $\delta_{t}^{V} = r_{t} + \gamma V(s_{t+1}) - V(s_{t})$.

It is important to set a sensible reward $R_{t}$. An intuitive design in aligning preference is to directly compute the score from the code snippets generated by the generator and ground truth directly. However, this does not take into consideration the features of the code generation task. Shorter code is generally easier to generate, resulting in higher reward. The corresponding relevant code is also shorter, leading the refactorer to conclude that generating shorter code increases the generator's performance, thus consistently producing more concise code snippets. During the reinforcement learning training process, the refactorer may generate tokens that could confuse the generator. To address this, we incorporate Kullback-Leibler divergence (KL) into the reward function to ensure that the output distribution of the policy model remains close to that of the original model. The final reward function is:
\begin{equation}
    R_{t} = \text{CodeBleu}(GT_{t},output_{t}) \cdot \sqrt{\text{len}(\text{tokenizer}(GT_{t}))} - \beta \text{KL}
\end{equation}

where $GT$ stands for ground truth, $output$ represents the output of the policy model, and $KL$ calculates the difference in output distribution between the policy model and the standard model. The standard model refers to the policy model before training.

\section{Experimental Setup}

\begin{table*}[ht]
\centering
\caption{Dataset Statistics. Code, NL, and Relevant refer to the average lengths after tokenization by CodeGPT. Relevant indicates the code retrieved by the retriever. Code tokens indicates the maximum length we set for the output generated by the refactorer.}
\begin{tabular}{c|c|C{1cm}|C{1cm}|C{1cm}|C{1cm}|C{1cm}|C{1.25cm}|c}
\hline
Dataset                        & Lang.  & Train  & Valid  & Test   & |Code| & |NL|  & |Relevant| & Code tokens \\ \hline
\multirow{2}{*}{CodeSearchNet} & java   & 454,451 & 15,328 & 26,909 & 152  & 58  & 449      & 168         \\
                               & python & 412,178 & 23,107 & 22,176 & 184  & 68  & 556      & 168         \\ \hline
ConCode                        & java   & 100,000 & 2,000  & 2,000  & 40   & 252 & 126      & 64          \\
\hline

\end{tabular}
\label{datasets}
\end{table*}

\subsection{Datasets}

To validate the effectiveness of RRG and simulate a RAG configuration, we utilize two datasets. We constructed an external knowledge base by merging the training, validation, and test sets from these datasets. To enhance data quality, we employed both javalang and Python's built-in AST (Abstract Syntax Tree) tools to filter out code with syntax errors. We then built a retrieval library and randomly selected 30,000 samples from the original training set to create a new, more refined training dataset. Although this significantly reduce the training set size, our model, after adopting SFT, still exhibited good performance, albeit slightly below the publicly available metrics.

The statistical information of the datasets and the code tokens for different dataset settings are shown in Table \ref{datasets}. It is worth noting that we trained the entire framework using only the sampled datasets without generating new data.

\begin{itemize}
    \item \textbf{CodeSearchNet} \cite{husain2019codesearchnet} is a dataset with 2 million annotated code pairs extracted from open source libraries hosted on GitHub. The dataset contains code and documentation for 6 programming languages. We utilize the python and java portions of the dataset.
    \item \textbf{ConCode} \cite{codegpt} is a highly specialized dataset for the java language, which incorporates natural language descriptions that cover not just the requirements but also the environment variables. Models using this dataset are expected to (a) exhibit a profound comprehension of the natural language description, translating it into environment variables, library API calls, and user-defined methods within the class, and (b) confidently ascertain the structure of the resulting code.
\end{itemize}

\subsection{Base Models}

To verify the robustness of our framework, we set up multiple retrievers and generators.

\textbf{Base models for Retriever.} We employ two types of retrievers: sparse retriever and dense retriever. For the sparse retriever, we use the BM25 algorithm to assess relevance. As for the dense retriever, we have utilized two different sets of semantic representation models: UAE-Large-V1 and mxbai-embed-large-v1. The similarity between the semantic representations is used to evaluate the similarity of the content. Since we are calculating the similarity between text semantics, we randomly selected two embedding models without the need to choose a model trained for code search. It is important to note that sparse retrieval often captures the lexical features of the text, while dense retrieval is more focused on capturing semantic features, resulting in different performance characteristics between the two approaches.

\textbf{Base models for Generator.} We set up multiple autoregressive language models with different scales of parameters and different pre-training methods:
\begin{itemize}
    \item GPT-2 \cite{gpt2} is a transformer model pre-trained on a large english corpus via self-supervised learning without requiring manual annotation. Widely applied across various natural language processing tasks, it has not undergone downstream task fine-tuning. We employ models of sizes 170M and 330M in our experiments.
    \item CodeGPT \cite{codegpt} shares similarities with the GPT-2 architecture but is trained on code corpora. Specifically, we utilize CodeGPT-small-java-adapted for java program and CodeGPT-small-py-adapted for python program.
    \item CodeGEN \cite{codegen} describes the process of generating code as a multi-round dialogue between the user and the system. This approach aims to significantly improve code generation performance by validating the effectiveness of the dialogic code generation paradigm. We use the CodeGEN-350M-mono version for experiments.
    \item PolyCoder \cite{xu2022polycoder} is based on the GPT-2 architecture and trained on a corpus of 12 programming languages, it performs better in C programming . We use 160M-sized models in our experiments.
    
\end{itemize}

\textbf{Base models for Refactorer.} For the refactor module, we adopted the 60M parameter version of CodeT5 \cite{codet5}, a unified pre-trained transformer model.

\subsection{Metrics}

In line with previous research, we utilize three key metrics to assess the quality of code generation in our method: Exact Match (EM), BLEU \cite{bleu}, and CodeBLEU \cite{ren2020codebleu}. EM determines whether the generated code is an exact match with the reference code. BLEU is an n-gram based metric that gauges the similarity between the generated code and the reference code, with scores ranging from 0 to 1. CodeBLEU, a variant of BLEU, is specifically designed for code generation tasks and takes into account the syntactic structure of the code.

When examining the impact of the different retrievers on performance of RRG, we introduce a new set of evaluation metrics, which includes cosine similarity and Levenshtein distance \cite{haldar2011levenshtein}. Both metrics aim to quantify the similarity between texts, albeit using different computational principles and applications. Cosine similarity is based on a vector space model, while Levenshtein distance calculates the minimum single-character edit distance between two text sequences.

\subsection{Implementation Details}

For dense retrieval, we utilize the FAISS \cite{douze2024faiss} database to store embeddings and similarity matching. For sparse retrieval, we implement it using the python library rank-bm25. We adopt a two-stage retrieval approach, setting Top-K$_{1}$ to 10 in the first stage and Top-K$_{2}$ to 3 in the second stage and utilize swifter to expedite pandas processing. In refactored fine-tuning phase, we set batch-size to 16, epoch to 10, and learning rate to 1e-5. During the preference-aware tuning phase of training, we also set the learning rate to 1e-5, batch-size to 16, and train for a single epoch using the first 5k entries in the dataset, with the KL scatter weight set to 0.5. The block-size of the generator and the refactorer is set to 512, and all experiments are conducted on four NVIDIA GeForce RTX 3090s.

\section{Experimental Results}

\subsection{Research Questions}

In the evaluation, we focus on the following five research questions:
\begin{itemize}
    \item \textbf{RQ1:} Does the RRG framework improve the effectiveness over RAG?
    \item \textbf{RQ2:} What is the effect of various retrievers on the performance of the RRG framework?
    \item \textbf{RQ3:} How do code tokens of varying lengths influence the RRG's performance?
    \item \textbf{RQ4:} What is the impact of the preference-aware tuning stage on RRG?   
    \item \textbf{RQ5:} How is the generalization performance of the frozen refactorer?
\end{itemize}

To investigate RQ1, we utilize two distinct datasets to simulate various scenarios. The ConCode dataset is characterized by relatively short code snippets with high similarity, introducing significant noise to the generator. In the RAG approach, all retrieved information is directly provided to the generator, whereas in the RRG framework, the generator solely refers to the output generated by the refactorer. Conversely, the CodeSearchNet dataset comprises longer and more intricate code, frequently exceeding the contextual window prescribed by the model. This complexity poses a challenge for the generator to produce precise outputs. Hence, we restrict the output size of both the retriever in RAG and the refactorer in RRG to a uniform length which is shown as code tokens in Table \ref{datasets}. The objective of these experiments is to affirm the efficacy of the RRG framework in enhancing the RAG model under diverse conditions.

Due to limited resources, all subsequent experiments for RQ2-RQ4 are conducted on both the ConCode and CodeSearchNet-java datasets. For RQ2, we configure multiple retriever settings to assess the robustness of RRG against variations in the retriever. Regarding RQ3, we systematically adjust the maximum length of code tokens in the refactorer's output and conduct extensive parameter experiments to scrutinize how diverse lengths influence RRG's performance. As for RQ4, we further delve into the significance of preference-aware tuning within RRG. By comparing experimental results across different datasets, generator models, and code token length settings, we thoroughly examine the impact of this training strategy on overall performance. For RQ5, in order to study the generalizability of the frozen refactorer, we directly transferred the trained refactorer to another different combination to observe the effects.

\subsection{Does the RRG framework improve the effectiveness over RAG?}

\begin{table*}[ht]
\centering
\caption{Comparing the performance of different generators in RRG and RAG frameworks. The results are based on ConCode and CodeSearchNet. The refactorer is based on CodeT5-small. The retriever employs a dual-stage retrieval that incorporates UAE and BM25.}
\begin{tabular}{c|c|C{1.2cm}C{1.2cm}C{1.2cm}C{1.2cm}C{1.2cm}C{1.2cm}C{1.2cm}C{1.2cm}C{1.2cm}}
\toprule
 &
   &
  \multicolumn{3}{c}{ConCode} &
  \multicolumn{3}{c}{CodeSearchNet-java} &
  \multicolumn{3}{c}{CodeSearchNet-python} \\ \cline{3-11} 
\multirow{-2}{*}{Base model} &
  \multirow{-2}{*}{Method} &
  EM &
  BLEU &
  CodeBLEU &
  EM &
  BLEU &
  CodeBLEU &
  EM &
  BLEU &
  CodeBLEU \\ \midrule
 &
  SFT &
  15.40 &
  27.42 &
  47.54 &
  4.10 &
  23.68 &
  35.32 &
  0.350 &
  10.14 &
  22.27 \\
 &
  \cellcolor[HTML]{EFF0F1}RAG &
  \cellcolor[HTML]{EFF0F1}16.75 &
  \cellcolor[HTML]{EFF0F1}37.08 &
  \cellcolor[HTML]{EFF0F1}52.86 &
  \cellcolor[HTML]{EFF0F1}6.65 &
  \cellcolor[HTML]{EFF0F1}35.07 &
  \cellcolor[HTML]{EFF0F1}45.57 &
  \cellcolor[HTML]{EFF0F1}1.0 &
  \cellcolor[HTML]{EFF0F1}17.33 &
  \cellcolor[HTML]{EFF0F1}28.94 \\
\multirow{-3}{*}{GPT2-s} &
  \cellcolor[HTML]{F0F4FF}RRG &
  \cellcolor[HTML]{F0F4FF}\textbf{19.00} &
  \cellcolor[HTML]{F0F4FF}\textbf{38.32} &
  \cellcolor[HTML]{F0F4FF}\textbf{54.52} &
  \cellcolor[HTML]{F0F4FF}\textbf{7.05} &
  \cellcolor[HTML]{F0F4FF}\textbf{37.61} &
  \cellcolor[HTML]{F0F4FF}\textbf{47.31} &
  \cellcolor[HTML]{F0F4FF}\textbf{1.15} &
  \cellcolor[HTML]{F0F4FF}\textbf{19.42} &
  \cellcolor[HTML]{F0F4FF}\textbf{30.92} \\ \midrule
 &
  SFT &
  15.67 &
  28.60 &
  47.99 &
  4.03 &
  24.13 &
  35.54 &
  0.4 &
  11.63 &
  23.82 \\
 &
  \cellcolor[HTML]{EFF0F1}RAG &
  \cellcolor[HTML]{EFF0F1}17.80 &
  \cellcolor[HTML]{EFF0F1}37.6 &
  \cellcolor[HTML]{EFF0F1}53.77 &
  \cellcolor[HTML]{EFF0F1}7.3 &
  \cellcolor[HTML]{EFF0F1}34.53 &
  \cellcolor[HTML]{EFF0F1}45.53 &
  \cellcolor[HTML]{EFF0F1}1.40 &
  \cellcolor[HTML]{EFF0F1}18.82 &
  \cellcolor[HTML]{EFF0F1}30.10 \\
\multirow{-3}{*}{PolyCoder} &
  \cellcolor[HTML]{F0F4FF}RRG &
  \cellcolor[HTML]{F0F4FF}\textbf{21.41} &
  \cellcolor[HTML]{F0F4FF}\textbf{39.26} &
  \cellcolor[HTML]{F0F4FF}\textbf{55.67} &
  \cellcolor[HTML]{F0F4FF}\textbf{8.0} &
  \cellcolor[HTML]{F0F4FF}\textbf{35.50} &
  \cellcolor[HTML]{F0F4FF}\textbf{45.77} &
  \cellcolor[HTML]{F0F4FF}\textbf{1.40} &
  \cellcolor[HTML]{F0F4FF}\textbf{19.63} &
  \cellcolor[HTML]{F0F4FF}\textbf{31.25} \\ \midrule
 &
  SFT &
  11.40 &
  26.82 &
  45.44 &
  3.9 &
  24.44 &
  35.71 &
  0.35 &
  10.45 &
  24.32 \\
 &
  \cellcolor[HTML]{EFF0F1}RAG &
  \cellcolor[HTML]{EFF0F1}11.85 &
  \cellcolor[HTML]{EFF0F1}31.73 &
  \cellcolor[HTML]{EFF0F1}49.74 &
  \cellcolor[HTML]{EFF0F1}\textbf{6.9} &
  \cellcolor[HTML]{EFF0F1}34.42 &
  \cellcolor[HTML]{EFF0F1}\textbf{45.04} &
  \cellcolor[HTML]{EFF0F1}\textbf{1.25} &
  \cellcolor[HTML]{EFF0F1}18.96 &
  \cellcolor[HTML]{EFF0F1}30.15 \\
\multirow{-3}{*}{CodeGEN} &
  \cellcolor[HTML]{F0F4FF}RRG &
  \cellcolor[HTML]{F0F4FF}\textbf{15.11} &
  \cellcolor[HTML]{F0F4FF}\textbf{33.76} &
  \cellcolor[HTML]{F0F4FF}\textbf{51.26} &
  \cellcolor[HTML]{F0F4FF}6.55 &
  \cellcolor[HTML]{F0F4FF}34.69 &
  \cellcolor[HTML]{F0F4FF}44.71 &
  \cellcolor[HTML]{F0F4FF}1.19 &
  \cellcolor[HTML]{F0F4FF}\textbf{19.37} &
  \cellcolor[HTML]{F0F4FF}\textbf{30.22} \\ \midrule
 &
  SFT &
  16.45 &
  29.00 &
  48.36 &
  4.8 &
  25.11 &
  36.55 &
  0.54 &
  11.33 &
  23.16 \\
 &
  \cellcolor[HTML]{EFF0F1}RAG &
  \cellcolor[HTML]{EFF0F1}17.55 &
  \cellcolor[HTML]{EFF0F1}36.77 &
  \cellcolor[HTML]{EFF0F1}53.05 &
  \cellcolor[HTML]{EFF0F1}\textbf{7.45} &
  \cellcolor[HTML]{EFF0F1}33.80 &
  \cellcolor[HTML]{EFF0F1}45.00 &
  \cellcolor[HTML]{EFF0F1}\textbf{0.95} &
  \cellcolor[HTML]{EFF0F1}17.83 &
  \cellcolor[HTML]{EFF0F1}29.22 \\
\multirow{-3}{*}{GPT2-m} &
  \cellcolor[HTML]{F0F4FF}RRG &
  \cellcolor[HTML]{F0F4FF}\textbf{20.86} &
  \cellcolor[HTML]{F0F4FF}\textbf{40.22} &
  \cellcolor[HTML]{F0F4FF}\textbf{55.55} &
  \cellcolor[HTML]{F0F4FF}7.25 &
  \cellcolor[HTML]{F0F4FF}\textbf{35.07} &
  \cellcolor[HTML]{F0F4FF}\textbf{45.38} &
  \cellcolor[HTML]{F0F4FF}0.90 &
  \cellcolor[HTML]{F0F4FF}\textbf{18.36} &
  \cellcolor[HTML]{F0F4FF}\textbf{30.04} \\ \midrule
 &
  SFT &
  15.85 &
  28.58 &
  48.36 &
  4.40 &
  24.01 &
  36.12 &
  0.4 &
  11.85 &
  23.65 \\
 &
  \cellcolor[HTML]{EFF0F1}RAG &
  \cellcolor[HTML]{EFF0F1}17.40 &
  \cellcolor[HTML]{EFF0F1}36.85 &
  \cellcolor[HTML]{EFF0F1}53.19 &
  \cellcolor[HTML]{EFF0F1}7.3 &
  \cellcolor[HTML]{EFF0F1}34.61 &
  \cellcolor[HTML]{EFF0F1}45.64 &
  \cellcolor[HTML]{EFF0F1}1.05 &
  \cellcolor[HTML]{EFF0F1}18.12 &
  \cellcolor[HTML]{EFF0F1}29.46 \\
\multirow{-3}{*}{CodeGPT} &
  \cellcolor[HTML]{F0F4FF}RRG &
  \cellcolor[HTML]{F0F4FF}\textbf{20.21} &
  \cellcolor[HTML]{F0F4FF}\textbf{39.40} &
  \cellcolor[HTML]{F0F4FF}\textbf{55.49} &
  \cellcolor[HTML]{F0F4FF}\textbf{8.20} &
  \cellcolor[HTML]{F0F4FF}\textbf{37.89} &
  \cellcolor[HTML]{F0F4FF}\textbf{47.63} &
  \cellcolor[HTML]{F0F4FF}\textbf{1.35} &
  \cellcolor[HTML]{F0F4FF}\textbf{20.50} &
  \cellcolor[HTML]{F0F4FF}\textbf{31.36} \\
\bottomrule
\end{tabular}

\label{rq1}
\end{table*}

To answer the RQ1, we evaluate the effectiveness of different generator models in code generation under various frameworks, and the experimental results are presented in Table \ref{rq1}.

RAG vs. SFT: Providing the generator with relevant code during both training and inference phases led to performance gains. This suggests that supplying the generator with contextual or guiding information effectively reduces the search space and allows the generator to focus on generating code relevant to the given task. The enhancement is more pronounced on the CodeSearchNet.

RRG vs. RAG: Overall, our proposed method RRG shows improvements in various metrics on both datasets while reducing the generator's cost. On the ConCode dataset, guiding the generator with the output from the refactorer leads to significant improvements across all metrics. Specifically, on the ConCode dataset, the EM metric increased by up to 28\%, the BLEU metric improved by 9\%, and CodeBLEU rose by 4.7\%. On the CodeSearchNet dataset, the EM metric increased by a maximum of 28\%, BLEU rose by up to 13\%, and CodeBLEU improved by a maximum of 6.8\%. This indicates that RRG exhibits good robustness and can enhance the performance of any generator on downstream tasks.

In the ConCode, there is a high degree of code similarity but significant differences in the implemented functions. The generator encounters more noise in the relevant codes retrieved by the retriever, which can mislead the generator, increase inference costs, and cause it to lost in middle. Directly inputting these relevant codes into the generator does not yield the best results. In RRG, removing redundant information and generating shorter, more efficient, and model-friendly code through refactorings significantly improves the generator's effectiveness.

The diversity of the CodeSearchNet is significantly higher than that of ConCode. When dealing with complex code, the refactorer proves capable of generating model-preferred formats. Remarkably, within the same context length, the refactored code demonstrates superior quality, thereby boosting the generator's overall effectiveness.

We observe that the CodeGEN performed worse on various datasets compared to models with fewer parameters. Comparing the results of GPT2-m and GPT2-s illustrates that models with larger parameters tend to achieve better performance. Therefore, the unusual performance of the CodeGEN may be attributed to its pre-training methodology, where the discrepancy between the pre-training and fine-tuning tasks has contributed to its inferior performance. Nonetheless, RRG still enhances the effectiveness of the CodeGEN.

\mdfsetup{linewidth=2pt,linecolor=gray,backgroundcolor=gray!20,roundcorner=5pt}
\begin{mdframed}[skipabove=5pt,innertopmargin=5pt,innerbottommargin=5pt,innerleftmargin=3pt,innerrightmargin=3pt]

\textbf{ANS to RQ1:} In most configurations, RRG can significantly enhance the performance of the generator and reduce its inference costs, demonstrating the robustness and effectiveness of RRG.
\end{mdframed}

\subsection{What is the effect of various retrievers on the performance of the RRG framework?}

\begin{table}[]
\centering
\caption{The effectiveness of various retrievers and how the generator performs with each of them. The generator is based on CodeGPT-java.}
\begin{tabular}{c|c|cccc}
\toprule
\multirow{2}{*}{Dataset}  & \multirow{2}{*}{Retriever} & \multicolumn{2}{c}{Retrieval} & \multicolumn{2}{c}{Generation} \\ \cline{3-6} 
                         &       & LS              & Cos similarity & EM             & BLEU  \\ \hline
\multirow{3}{*}{ConCode} & BM25  & 98.90           & 0.35           & 18.55          & 35.43 \\
                         & UAE   & 95.71           & 0.38           & \textbf{19.11} & 36.86 \\
                         & Mxbai & \textbf{94.98}  & \textbf{0.39}  & 17.85          & \textbf{37.24} \\ \hline
\multirow{3}{*}{CSN-java} & BM25                       & 293.52         & 0.30         & \textbf{8.05} & \textbf{35.37} \\
                         & UAE   & 288.69          & 0.32           & 7.2            & 35.20 \\
                         & Mxbai & \textbf{287.88} & \textbf{0.33}  & 7.05           & 33.08 \\
\bottomrule

\end{tabular}

\label{gaps-exp}

\end{table}

\begin{figure*}[h]
    \centering
    \begin{subfigure}[b]{0.48\textwidth}
        \centering
        \includegraphics[width=\textwidth]{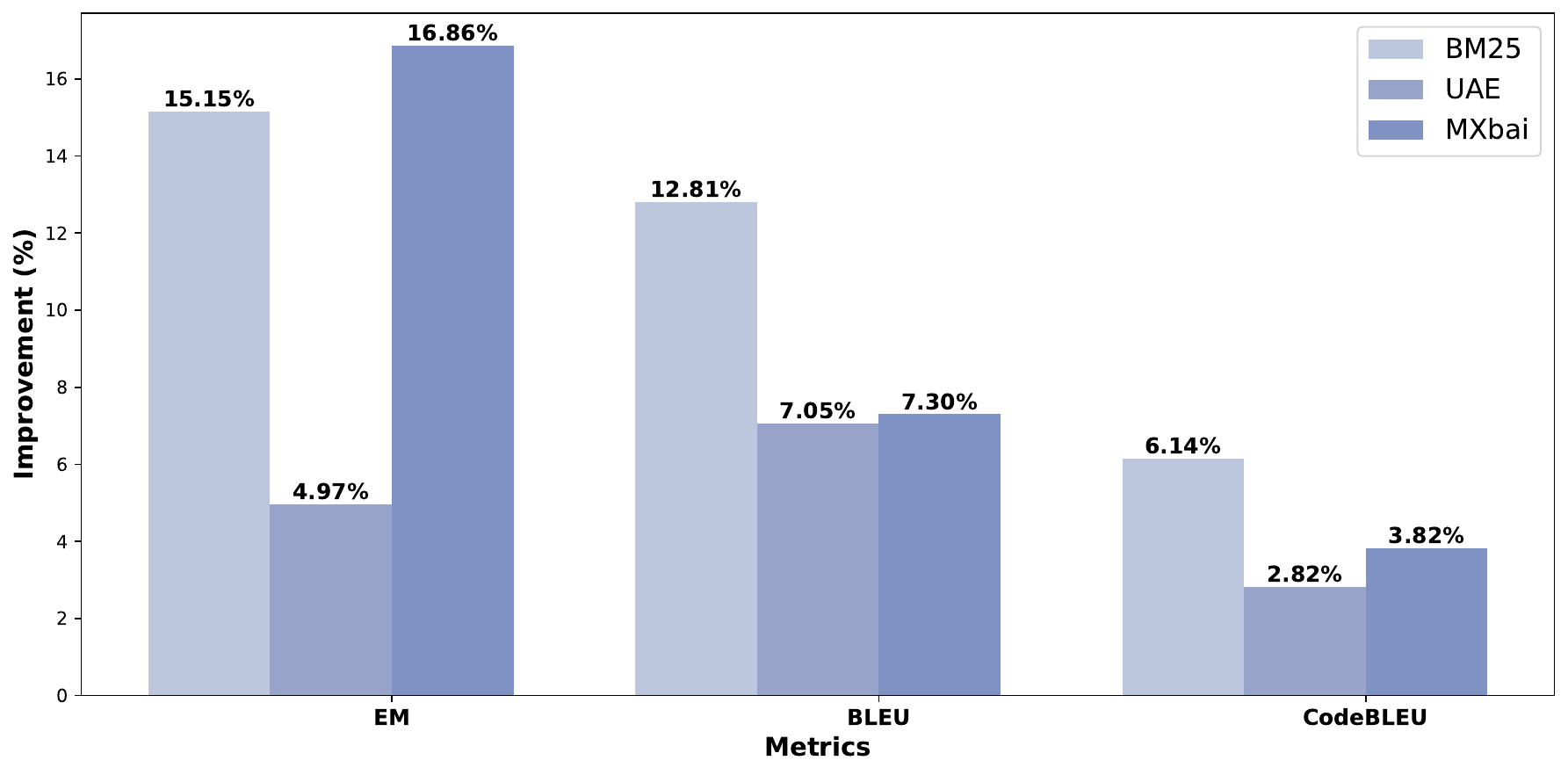}
        \caption{The improvements on the ConCode dataset.}
    \end{subfigure}
    \hfill
    \begin{subfigure}[b]{0.48\textwidth}
        \centering
        \includegraphics[width=\textwidth]{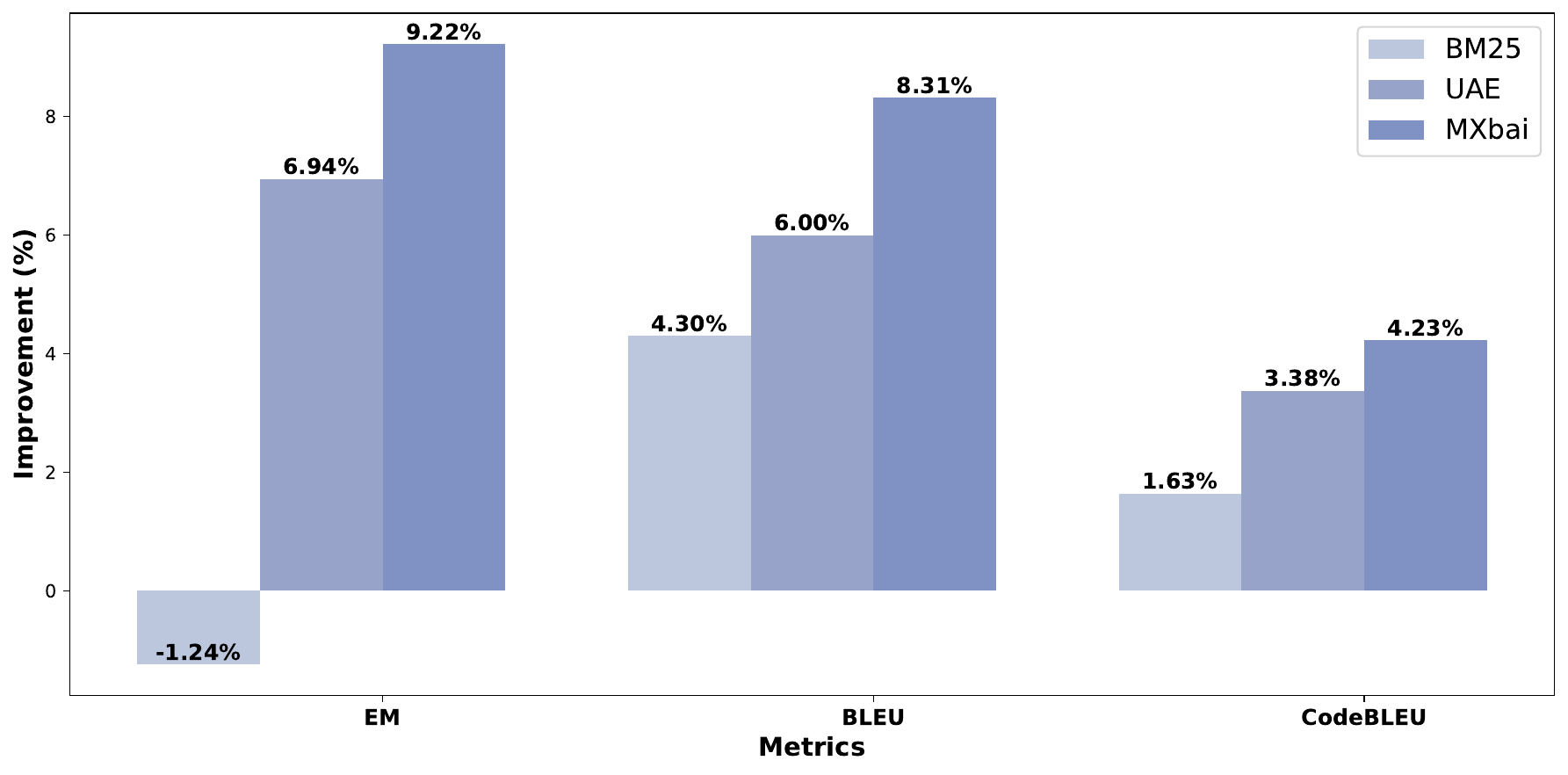}
        \caption{The improvements on the  CSN-java dataset.}
    \end{subfigure}
    
    \caption{Shows the improvements brought by the RRG framework when configured with different retrievers. The generator based on CodeGPT-java.}
    \label{impact-retrieve}
\end{figure*}

To answer RQ2, we compare the improvements of RRG on generator performance under different retriever settings, including sparse retrieval BM25 and two semantic retrieval models. Table \ref{gaps-exp} shows the retrieval results of these retrievers in naive RAG and their generator output. Fig. \ref{impact-retrieve} shows the improvement of the generator by the RRG framework.

On the one hand, the experimental results show that the BM25 algorithm retrieves the worst results. Because natural language descriptions contain some useless words, recalling the target code based on lexical features alone does not meet human preferences. On the other hand, the other vector retrievers differ in the number of parameters and pre-training methods, leading to differences in retrieval results. It is worth noting that the performance of the generator is not always directly correlated with the performance of the retriever. Although mxbai, which had the best results on each dataset, is used to guide the generator, the optimum generation results are not obtained. Similarly, BM25 does not lead to the worst generator results. For example, on the ConCode dataset, the generator produces the worst generation accuracy when using mxbai, whereas on the CSN-java dataset, BM25 achieve the best results on both metrics. This suggests that there is indeed a preference gap between the retriever and the generator.

Overall, as shown in Fig. \ref{impact-retrieve}, the RRG framework can bridge the preference gap between the retriever and the generator by inserting a refactorer between them. On the ConCode dataset, RRG can improve the EM metric by up to 16.86\%, BLEU by up to 12.81\%, and CodeBLEU by up to 6.14\%. On the CSN-java dataset, it can improve EM by up to 4.30\%, BLEU by 6.94\%, and CodeBLEU by 9.22\%. This demonstrates the robustness of RRG in the face of different retrievers.
However, on the CSN-java dataset, we notice a decrease in the EM metric when using BM25 as the retriever, and the other two metrics do not improve as significantly as with other retrievers. We deem that this is caused by the preference between the retriever and the generator. As shown in Table \ref{gaps-exp}, on the CSN-java dataset, the BM25 algorithm retrieves the relevant code that can best guide the generator. The preference of the two is relatively the same. If we then introduce refactoring for the alignment, the opposite effect may be triggered. For other retrievers, the gap between the two is large, and the use of refactoring can significantly improve the final performance of the generator.

\mdfsetup{linewidth=2pt,linecolor=gray,backgroundcolor=gray!20,roundcorner=5pt}
\begin{mdframed}[skipabove=5pt,innertopmargin=5pt,innerbottommargin=5pt,innerleftmargin=3pt,innerrightmargin=3pt]

\textbf{ANS to RQ2:} The RRG framework can perform well when using different base models as retrievers. Moreover, in code generation tasks, there is indeed a preference gap between retrievers and generators, and RRG can bridge the gap.
\end{mdframed}

\subsection{How do code tokens of varying lengths influence the RRG's performance?}

\begin{figure*}[h]
    \centering

    \includegraphics[scale = 0.48]{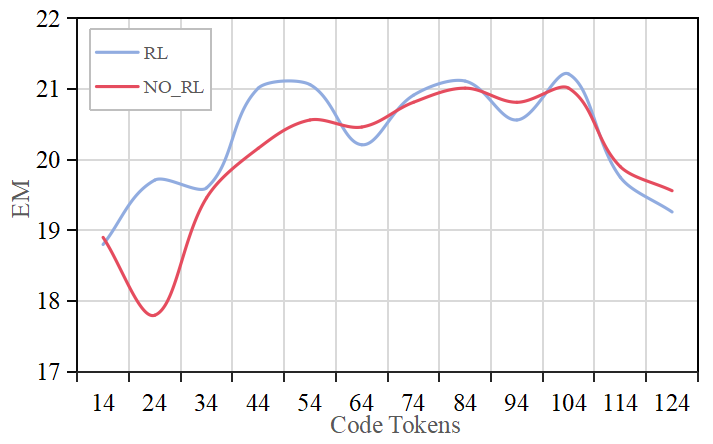}
    \includegraphics[scale = 0.47]{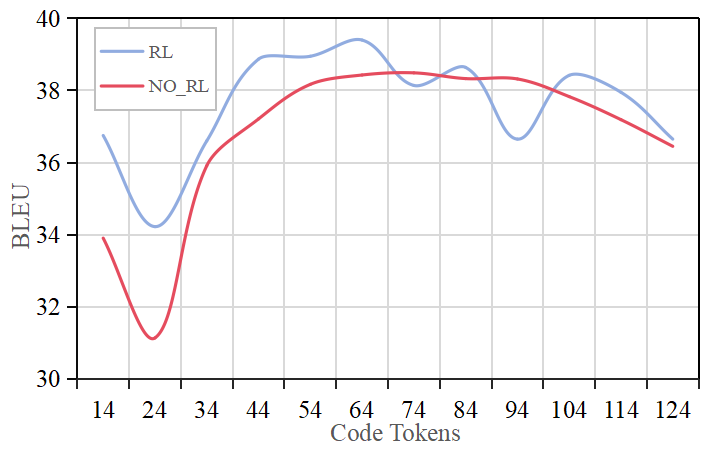}
    \includegraphics[scale = 0.47]{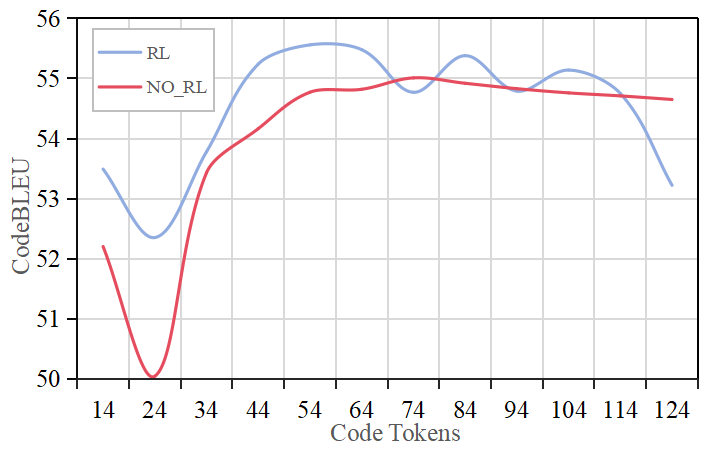}
    \begin{center}
        \textbf{(a) Result on the ConCode dataset.}
    \end{center}
    \hfill
    \includegraphics[scale = 0.48]{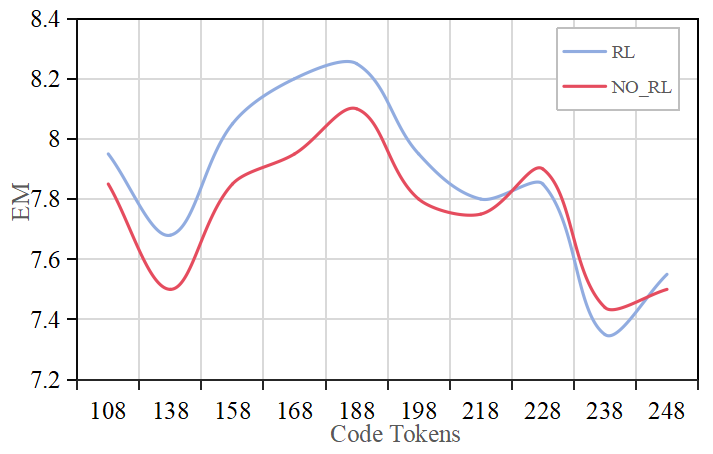}
    \includegraphics[scale = 0.47]{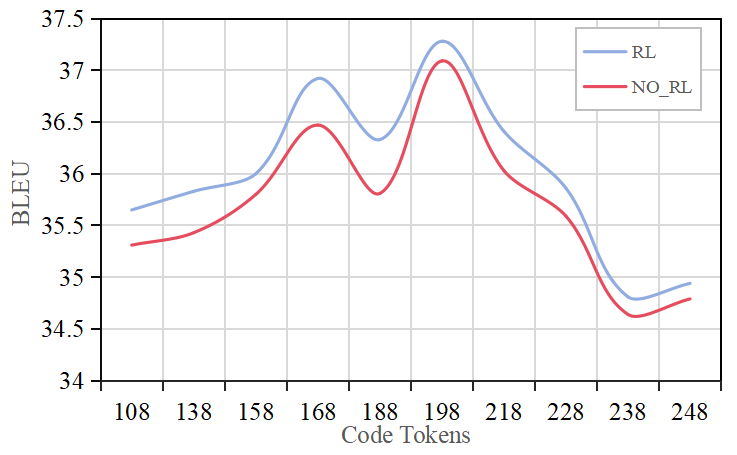}
    \includegraphics[scale = 0.47]{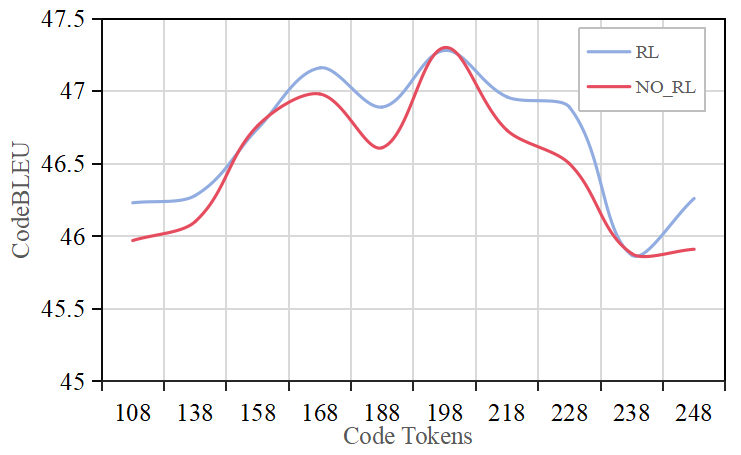}
    \begin{center}
        \textbf{(b) Result on the CSN-java dataset.}
    \end{center}

    \caption{Shows the performance of RL and NO-RL in different code tokens settings}
    
    \label{rq3}
    
\end{figure*}

To answer RQ3, in this set of experiments, we investigated the effect of different code token settings on the code generation performance of RRG. Specifically, we performed a comparative analysis by limiting the maximum number of output tokens of refactorer. The experimental results are shown in Fig. \ref{rq3}. It is demonstrated that the length of code tokens has a significant impact on the code generation performance because different numbers of code tokens represent the amount of information passed to the generator. This characteristic directly affects the performance of the model in generating code.
Notably, we observe that the results show a trend similar to a normal distribution. When the number of code tokens is set low, the amount of information passed to the model is low, leading to relatively poor results. As the number of tokens increases, the performance gradually improves. However, it is worth noting that when the number of tokens reaches a certain threshold, too much information may lead to a redundancy phenomenon, which in turn reduces the code generation performance.
In the ConCode dataset, the effect of different information lengths on the model performance is significant, with a 12.02 difference between the extremes of accuracy. In the CodeSearchNet-java dataset, this difference reaches 30\%. This suggests that the relationship between the amount of information and generation effectiveness needs to be carefully weighed in practical applications.

\mdfsetup{linewidth=2pt,linecolor=gray,backgroundcolor=gray!20,roundcorner=5pt}
\begin{mdframed}[skipabove=5pt,innertopmargin=5pt,innerbottommargin=5pt,innerleftmargin=3pt,innerrightmargin=3pt]

\textbf{ANS to RQ3:} The setting of different code tokens has a significant impact on the code generation performance of RRG, with this difference reaching 12.02\% and 30\% on two datasets, respectively.
\end{mdframed}

\subsection{What is the impact of the preference-aware tuning stage on RRG?}

\begin{table}[h]
\centering
\caption{Ablation Study on ConCode and CodeSearchNet-java datasets. The retriever employs a dual-stage retrieval that incorporates UAE and BM25.}
\begin{tabular}{ccccc}
\hline
\multicolumn{5}{c}{ConCode}                                                                                                           \\ \hline
\multicolumn{1}{c|}{Model Name}                 & \multicolumn{1}{c|}{Method}      & EM             & BLEU           & CodeBLEU       \\ \hline
\multicolumn{1}{c|}{\multirow{2}{*}{CodeGPT}}   & \multicolumn{1}{c|}{RRG w/o RL}   & 19.65          & 38.24          & \textbf{55.12} \\
\multicolumn{1}{c|}{}                           & \multicolumn{1}{c|}{RRG}        & \textbf{19.71} & \textbf{38.58} & 54.97          \\ \hline
\multicolumn{1}{c|}{\multirow{2}{*}{PolyCoder}} & \multicolumn{1}{c|}{RRG w/o RL}   & 20.76          & \textbf{40.40} & \textbf{55.77} \\
\multicolumn{1}{c|}{}                           & \multicolumn{1}{c|}{RRG}        & \textbf{21.41} & 39.26          & 55.67          \\ \hline
\multicolumn{1}{c|}{\multirow{2}{*}{CodeGEN}}   & \multicolumn{1}{c|}{RRG w/o RL} & 15.01          & 33.03          & 50.97          \\
\multicolumn{1}{c|}{}                           & \multicolumn{1}{c|}{RRG} & \textbf{15.11} & \textbf{33.76} & \textbf{51.26} \\ \hline
\multicolumn{5}{c}{CodeSearchNet-java}                                                                                                \\ \hline
\multicolumn{1}{c|}{Model Name}                 & \multicolumn{1}{c|}{Method}      & EM             & BLEU           & CodeBLEU       \\ \hline
\multicolumn{1}{c|}{\multirow{2}{*}{CodeGPT}}   & \multicolumn{1}{c|}{RRG w/o RL} & 7.44           & 37.04          & 47.02          \\
\multicolumn{1}{c|}{}                           & \multicolumn{1}{c|}{RRG}        & \textbf{7.7}   & \textbf{37.73} & \textbf{47.41} \\ \hline
\multicolumn{1}{c|}{\multirow{2}{*}{PolyCoder}} & \multicolumn{1}{c|}{RRG w/o RL} & 7.85           & 35.18          & 45.52          \\
\multicolumn{1}{c|}{}                           & \multicolumn{1}{c|}{RRG}        & \textbf{8.0}   & \textbf{35.50} & \textbf{45.77} \\ \hline
\multicolumn{1}{c|}{\multirow{2}{*}{CodeGEN}}   & \multicolumn{1}{c|}{RRG w/o RL} & 6.55           & 34.69          & 44.71          \\
\multicolumn{1}{c|}{}                           & \multicolumn{1}{c|}{RRG}        & \textbf{6.9}   & \textbf{34.86} & \textbf{44.94} \\ \hline
\end{tabular}

\label{ablation}

\end{table}

To answer the RQ4, we investigate the impact of the Preference-aware Tuning phase on the effectiveness of RRG. We conducted an ablation study. The experimental results are presented in Table \ref{ablation} and Fig. \ref{impact-retrieve}. Overall, removing the preference-aware tuning phase resulted in poorer performance of RRG across different models, datasets, and code tokens configurations. This decline is attributed to the fact that after the SFT phase, the output of the refactorer is aligned with human preferences without considering the preferences of the generator. In contrast, when the Preference-aware Tuning is incorporated during training, the refactorer's output is directly aligned with the generator's preferences, helping RRG generate more effective code tokens.

Furthermore, on the CodeSearchNet-java dataset, we notice that preference-aware tuning consistently improved the performance across all models. However, it is not the case for the ConCode dataset, suggesting that this phase is better suited for more complex datasets. 


\mdfsetup{linewidth=2pt,linecolor=gray,backgroundcolor=gray!20,roundcorner=5pt}
\begin{mdframed}[skipabove=5pt,innertopmargin=5pt,innerbottommargin=5pt,innerleftmargin=3pt,innerrightmargin=3pt]

\textbf{ANS to RQ4:} The preference-aware tuning stage is crucial, as it further enhances the effectiveness of the RRG framework by considering the preference of the generator. Moreover, it exhibits good generalization, potentially yielding better results on complex datasets.
\end{mdframed}

\subsection{How is the generalization performance of the frozen refactorer ?}


\begin{table}[]
\centering
\caption{Generalization experiment. The original combinationsin the table mean where the frozen refactorer was trained. And we directly applied it to other combinations on the ConCode.}
\begin{tabular}{c|ccc}
\hline
Combinations     & Retriever-Generator   & EM             & BLEU           \\ \hline
\rowcolor[HTML]{F0F4FF} 
original         & \textbf{mxbai-GPT2}   & \textbf{20.01} & \textbf{38.34} \\
change-retriever & BM25-GPT2             & 18.25          & 36.73          \\
change-generator & mxbai-CodeGPT         & 19.35          & 38.31          \\
change-both      & BM25-CodeGPT          & 20.41          & 39.11          \\ \hline
\rowcolor[HTML]{F0F4FF} 
original         & \textbf{BM25-CodeGPT} & \textbf{20.81} & \textbf{39.62} \\
change-retriever & mxbai-CodeGPT         & 19.25          & 37.73          \\
change-generator & BM25-GPT2             & 18.25          & 36.73          \\
change-both      & mxbai-GPT2            & 19.75          & 38.24          \\ \hline
\end{tabular}
\label{frozen}
\end{table}

We froze the trained refactorer and experimented with various retriever and generator combinations to evaluate the performance of RRG in code generation. The original combinations listed in the table \ref{frozen} represent the configurations where the refactorer was initially trained. Experimental results indicate that even when frozen, the refactorer aligns well with diverse retriever and generator combinations, demonstrating strong generalization capabilities.

However, a comparison between the mxbai-GPT2 and BM25-CodeGPT configurations reveals that applying a trained refactorer to a new combination results in less effective outcomes compared to using a refactorer specifically trained for that combination. This suggests that, while the refactorer exhibits generalization, specialized training for particular combinations still yields better performance.

Given the high costs associated with retraining models, the refactorer's generalization ability is particularly noteworthy. In practical applications, this means the refactorer can adapt to different requirements or scenarios by simply altering the retriever and generator combinations, maintaining robust performance while significantly reducing development and maintenance costs.

\mdfsetup{linewidth=2pt,linecolor=gray,backgroundcolor=gray!20,roundcorner=5pt}
\begin{mdframed}[skipabove=5pt,innertopmargin=5pt,innerbottommargin=5pt,innerleftmargin=3pt,innerrightmargin=3pt]

\textbf{ANS to RQ5:} The frozen refactorer demonstrates strong generalizability, suggesting its advantage as a modular, plug-and-play component. In practice, it can be directly transferred between different retrievers and generators without requiring fine-tuning.
\end{mdframed}

\subsection{Case Study}

\begin{figure*}
    \centering
    \includegraphics[scale=0.78]{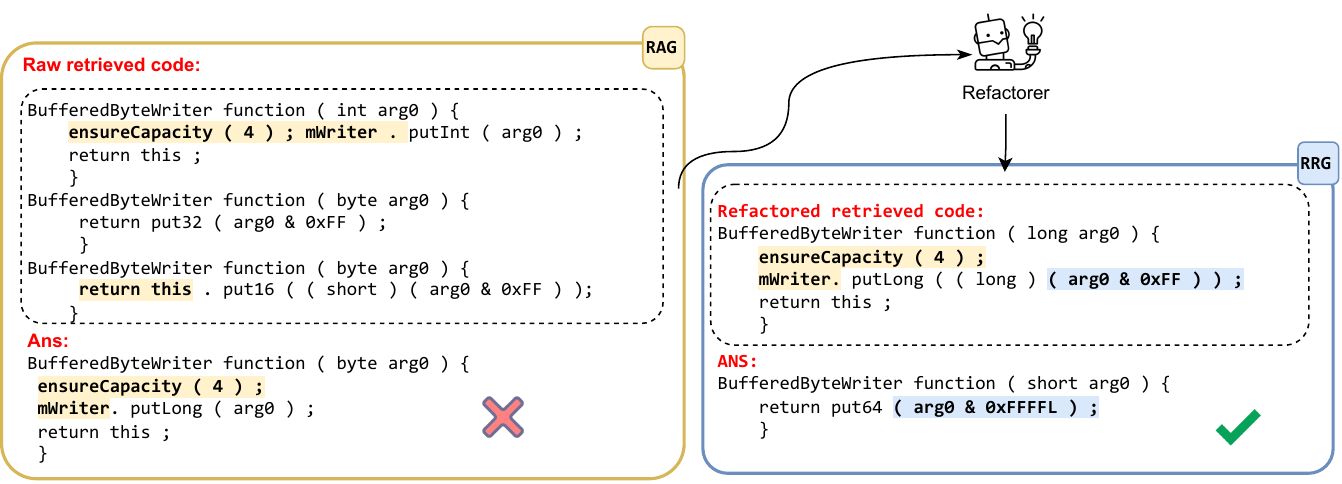}
    \caption{Case comparison. Redundant info and preference gaps led the generator to produce wrong answers from raw retrieved code, whereas it generated correctly from the refactored code. Highlighted tokens significantly impact code generation, whereas misleading tokens in the raw do not affect the generator in the refactored version.}
    \label{case-study1}
\end{figure*}

\begin{figure}[h]
    \centering
    \includegraphics[scale=0.73]{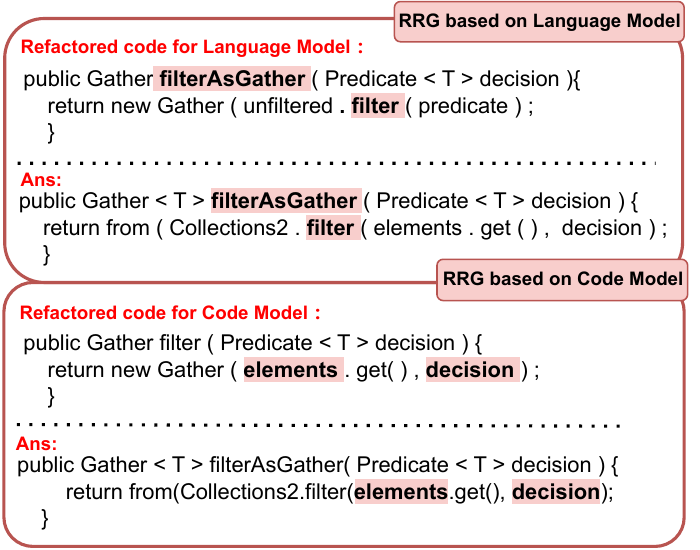}
    \caption{The preference-guided refactorer will provide different code to different generators. The highlight represents the tokens that play a significant role in guiding the model's generation. This difference is referred to as the preferences of different generators.}
    \label{case-study2}
\end{figure}

We illustrate an example of RRG's effectiveness in Fig. \ref{case-study1}. When the relevant code provided by the retriever contains redundant information, the generator may be misled and produce incorrect code. The generated code closely resembles the retrieved code and is heavily influenced by the top-1 code. However, when using refactored code, the generator can produce correct code. Due to certain model preferences, such as a tendency to favor certain token locations or repetitions, the generator may focus more on the irrelevant tokens, resulting in incorrect output. In contrast, the refactored code retains the tokens that mislead the generator in the raw but is shorter and contains less redundant information. This higher-quality context allows the generator to produce correct answers more easily and avoid being misled. We show an example of preferences between different models in Fig. \ref{case-study2}. Language models are based on GPT2, while code models are based on CodeGPT. We found that the refactorer provides different codes for different generators. For GPT2, the $filter$ token plays a significant role in the refactored code, while CodeGPT is $elements$ token and $decision$ token. We interpret this phenomenon as a result of inherent preference differences among various LLMs. Due to different internal knowledge, generating a piece of code requires different codes to be provided.

\section{Threats to Validity}

\textbf{Issue of model scale.} The generators in our experiments are mainly based on a number of small code models with a relatively low number of parameters. But our approach can still improve the effectiveness of these models with relatively low capacity. However, we also realize that this may not reflect the applicability of our framework when dealing with larger scale models. We compared models with different numbers of parameters in our experiments, in which models with a larger number of parameters can indeed achieve better results in downstream tasks. 
The same holds true for scaling-law in our framework. Furthermore, our reinforcement learning phase also applies to black-box models that do not require access to their internal parameters.


\textbf{Framework setup of the baseline.} In our experiments, we mainly compared the improvements that RRG brought to RAG and did not compare it with other RACG methods. However, this does not indicate a fundamental flaw in our research. In fact, our refactorer plugin is designed with a high degree of modularity and plug-and-play capability, meaning it can be easily integrated into any existing framework without significant changes to the original system. When RRG is combined with other methods, it may produce even more impressive results.

\section{Conclusion and Future Works}

In this paper, we introduce RRG, a novel retrieval-augmented code generation framework, designed to tackle redundant information and bridge the preference gap between retrievers and generators. The preference gap arises from the mismatch between retrieved content and the generator's actual needs. By integrating a refactor module, we harness both retrieved and internal knowledge to supply the generator with optimized code snippets. Our experiments reveal that RRG significantly enhances code generation with strong generalization. 

Furthermore, we uncover that retrievers and generators exhibit distinct preferences in RACG, and RRG effectively aligns these. Exploring and utilizing specific generator preferences offers vast potential, yet selecting the most beneficial documents based on these remains an open challenge. Currently, we treat model preferences as a black box. However, we aspire to enhance interpretability by elucidating these preferences in the future. Meanwhile, we have also observed that filtering information through RRG can sometimes lead to worse outcomes, highlighting the importance of an adaptive refactorer module. But inspired by soft prompts, perhaps we can achieve the same or even better results with fewer tokens. In the post-processing of retrieval-augmented code generation, there are still unknown territories worth exploring. 

\section{Acknowledgments}
This work is funded in part by 2022-JCJQ-LA-001-096,  the Shanghai Science, Technology Development Fund No. 22dz1200704, the National Natural Science Foundation of China Projects No. U1936213, 62272473, and the Science and Technology Innovation Program of Hunan Province No.2023RC1001.

\bibliographystyle{plain}
\bibliography{sample-base}

\begin{thebibliography}{10}

\bibitem{santacoder}
Loubna~Ben Allal, Raymond Li, Denis Kocetkov, Chenghao Mou, Christopher Akiki,
  Carlos~Munoz Ferrandis, Niklas Muennighoff, Mayank Mishra, Alex Gu, Manan
  Dey, et~al.
\newblock Santacoder: don't reach for the stars!
\newblock {\em arXiv preprint arXiv:2301.03988}, 2023.

\bibitem{self-rag}
Akari Asai, Zeqiu Wu, Yizhong Wang, Avirup Sil, and Hannaneh Hajishirzi.
\newblock Self-rag: Learning to retrieve, generate, and critique through
  self-reflection.
\newblock {\em arXiv preprint arXiv:2310.11511}, 2023.

\bibitem{token_elimination}
Moshe Berchansky, Peter Izsak, Avi Caciularu, Ido Dagan, and Moshe Wasserblat.
\newblock Optimizing retrieval-augmented reader models via token elimination.
\newblock {\em arXiv preprint arXiv:2310.13682}, 2023.

\bibitem{intro}
Sebastian Borgeaud, Arthur Mensch, Jordan Hoffmann, Trevor Cai, Eliza
  Rutherford, Katie Millican, George~Bm Van Den~Driessche, Jean-Baptiste
  Lespiau, Bogdan Damoc, Aidan Clark, et~al.
\newblock Improving language models by retrieving from trillions of tokens.
\newblock In {\em International conference on machine learning}, pages
  2206--2240. PMLR, 2022.

\bibitem{gpt3}
Tom Brown, Benjamin Mann, Nick Ryder, Melanie Subbiah, Jared~D Kaplan, Prafulla
  Dhariwal, Arvind Neelakantan, Pranav Shyam, Girish Sastry, Amanda Askell,
  et~al.
\newblock Language models are few-shot learners.
\newblock {\em Advances in neural information processing systems},
  33:1877--1901, 2020.

\bibitem{devlin2018bert}
Jacob Devlin, Ming-Wei Chang, Kenton Lee, and Kristina Toutanova.
\newblock Bert: Pre-training of deep bidirectional transformers for language
  understanding.
\newblock {\em arXiv preprint arXiv:1810.04805}, 2018.

\bibitem{douze2024faiss}
Matthijs Douze, Alexandr Guzhva, Chengqi Deng, Jeff Johnson, Gergely Szilvasy,
  Pierre-Emmanuel Mazaré, Maria Lomeli, Lucas Hosseini, and Hervé Jégou.
\newblock The faiss library.
\newblock 2024.

\bibitem{gong2024ast}
Linyuan Gong, Mostafa Elhoushi, and Alvin Cheung.
\newblock Ast-t5: Structure-aware pretraining for code generation and
  understanding.
\newblock {\em arXiv preprint arXiv:2401.03003}, 2024.

\bibitem{rag}
Kelvin Guu, Kenton Lee, Zora Tung, Panupong Pasupat, and Mingwei Chang.
\newblock Retrieval augmented language model pre-training.
\newblock In {\em International conference on machine learning}, pages
  3929--3938. PMLR, 2020.

\bibitem{haldar2011levenshtein}
Rishin Haldar and Debajyoti Mukhopadhyay.
\newblock Levenshtein distance technique in dictionary lookup methods: An
  improved approach.
\newblock {\em arXiv preprint arXiv:1101.1232}, 2011.

\bibitem{husain2019codesearchnet}
Hamel Husain, Ho-Hsiang Wu, Tiferet Gazit, Miltiadis Allamanis, and Marc
  Brockschmidt.
\newblock Codesearchnet challenge: Evaluating the state of semantic code
  search.
\newblock {\em arXiv preprint arXiv:1909.09436}, 2019.

\bibitem{jiang-etal-2023-llmlingua}
Huiqiang Jiang, Qianhui Wu, Chin-Yew Lin, Yuqing Yang, and Lili Qiu.
\newblock {LLML}ingua: Compressing prompts for accelerated inference of large
  language models.
\newblock In Houda Bouamor, Juan Pino, and Kalika Bali, editors, {\em
  Proceedings of the 2023 Conference on Empirical Methods in Natural Language
  Processing}, pages 13358--13376, Singapore, December 2023. Association for
  Computational Linguistics.

\bibitem{Active}
Zhengbao Jiang, Frank~F Xu, Luyu Gao, Zhiqing Sun, Qian Liu, Jane Dwivedi-Yu,
  Yiming Yang, Jamie Callan, and Graham Neubig.
\newblock Active retrieval augmented generation.
\newblock {\em arXiv preprint arXiv:2305.06983}, 2023.

\bibitem{long-tail}
Nikhil Kandpal, Haikang Deng, Adam Roberts, Eric Wallace, and Colin Raffel.
\newblock Large language models struggle to learn long-tail knowledge.
\newblock In {\em International Conference on Machine Learning}, pages
  15696--15707. PMLR, 2023.

\bibitem{bgm}
Zixuan Ke, Weize Kong, Cheng Li, Mingyang Zhang, Qiaozhu Mei, and Michael
  Bendersky.
\newblock Bridging the preference gap between retrievers and llms.
\newblock {\em arXiv preprint arXiv:2401.06954}, 2024.

\bibitem{rag-token1}
Urvashi Khandelwal, Omer Levy, Dan Jurafsky, Luke Zettlemoyer, and Mike Lewis.
\newblock Generalization through memorization: Nearest neighbor language
  models.
\newblock {\em arXiv preprint arXiv:1911.00172}, 2019.

\bibitem{rag-dy1}
Tian Lan, Deng Cai, Yan Wang, Heyan Huang, and Xian-Ling Mao.
\newblock Copy is all you need.
\newblock In {\em The Eleventh International Conference on Learning
  Representations}, 2023.

\bibitem{lail}
Jia Li, Ge~Li, Chongyang Tao, Huangzhao Zhang, Fang Liu, and Zhi Jin.
\newblock Large language model-aware in-context learning for code generation.
\newblock {\em arXiv preprint arXiv:2310.09748}, 2023.

\bibitem{webRAG}
Junyi Li, Tianyi Tang, Wayne~Xin Zhao, Jingyuan Wang, Jian-Yun Nie, and Ji-Rong
  Wen.
\newblock The web can be your oyster for improving large language models.
\newblock {\em arXiv preprint arXiv:2305.10998}, 2023.

\bibitem{li2023unlocking}
Yucheng Li.
\newblock Unlocking context constraints of llms: Enhancing context efficiency
  of llms with self-information-based content filtering.
\newblock {\em arXiv preprint arXiv:2304.12102}, 2023.

\bibitem{Less-is-More}
Chen Liang, Simiao Zuo, Qingru Zhang, Pengcheng He, Weizhu Chen, and Tuo Zhao.
\newblock Less is more: Task-aware layer-wise distillation for language model
  compression.
\newblock pages 20852--20867, 2022.

\bibitem{liang2024repofuse}
Ming Liang, Xiaoheng Xie, Gehao Zhang, Xunjin Zheng, Peng Di, Hongwei Chen,
  Chengpeng Wang, Gang Fan, et~al.
\newblock Repofuse: Repository-level code completion with fused dual context.
\newblock {\em arXiv preprint arXiv:2402.14323}, 2024.

\bibitem{radit}
Xi~Victoria Lin, Xilun Chen, Mingda Chen, Weijia Shi, Maria Lomeli, Rich James,
  Pedro Rodriguez, Jacob Kahn, Gergely Szilvasy, Mike Lewis, et~al.
\newblock Ra-dit: Retrieval-augmented dual instruction tuning.
\newblock {\em arXiv preprint arXiv:2310.01352}, 2023.

\bibitem{codegen4lib}
Mingwei Liu, Tianyong Yang, Yiling Lou, Xueying Du, Ying Wang, and Xin Peng.
\newblock Codegen4libs: A two-stage approach for library-oriented code
  generation.
\newblock In {\em 2023 38th IEEE/ACM International Conference on Automated
  Software Engineering (ASE)}, pages 434--445, 2023.

\bibitem{lost}
Nelson~F Liu, Kevin Lin, John Hewitt, Ashwin Paranjape, Michele Bevilacqua,
  Fabio Petroni, and Percy Liang.
\newblock Lost in the middle: How language models use long contexts.
\newblock {\em Transactions of the Association for Computational Linguistics},
  12:157--173, 2024.

\bibitem{webglm}
Xiao Liu, Hanyu Lai, Hao Yu, Yifan Xu, Aohan Zeng, Zhengxiao Du, Peng Zhang,
  Yuxiao Dong, and Jie Tang.
\newblock Webglm: Towards an efficient web-enhanced question answering system
  with human preferences.
\newblock In {\em Proceedings of the 29th ACM SIGKDD Conference on Knowledge
  Discovery and Data Mining}, pages 4549--4560, 2023.

\bibitem{codegpt}
Shuai Lu, Daya Guo, Shuo Ren, Junjie Huang, Alexey Svyatkovskiy, Ambrosio
  Blanco, Colin Clement, Dawn Drain, Daxin Jiang, Duyu Tang, et~al.
\newblock Codexglue: A machine learning benchmark dataset for code
  understanding and generation.
\newblock {\em arXiv preprint arXiv:2102.04664}, 2021.

\bibitem{rag-dy2}
Sewon Min, Weijia Shi, Mike Lewis, Xilun Chen, Wen-tau Yih, Hannaneh
  Hajishirzi, and Luke Zettlemoyer.
\newblock Nonparametric masked language modeling.
\newblock {\em arXiv preprint arXiv:2212.01349}, 2022.

\bibitem{codegen}
Erik Nijkamp, Bo~Pang, Hiroaki Hayashi, Lifu Tu, Huan Wang, Yingbo Zhou, Silvio
  Savarese, and Caiming Xiong.
\newblock Codegen: An open large language model for code with multi-turn
  program synthesis.
\newblock {\em arXiv preprint arXiv:2203.13474}, 2022.

\bibitem{pan2024llmlingua}
Zhuoshi Pan, Qianhui Wu, Huiqiang Jiang, Menglin Xia, Xufang Luo, Jue Zhang,
  Qingwei Lin, Victor R{\"u}hle, Yuqing Yang, Chin-Yew Lin, et~al.
\newblock Llmlingua-2: Data distillation for efficient and faithful
  task-agnostic prompt compression.
\newblock {\em arXiv preprint arXiv:2403.12968}, 2024.

\bibitem{bleu}
Kishore Papineni, Salim Roukos, Todd Ward, and Wei-Jing Zhu.
\newblock {B}leu: a method for automatic evaluation of machine translation.
\newblock In Pierre Isabelle, Eugene Charniak, and Dekang Lin, editors, {\em
  Proceedings of the 40th Annual Meeting of the Association for Computational
  Linguistics}, pages 311--318, Philadelphia, Pennsylvania, USA, July 2002.
  Association for Computational Linguistics.

\bibitem{recoder}
Md~Rizwan Parvez, Wasi~Uddin Ahmad, Saikat Chakraborty, Baishakhi Ray, and
  Kai-Wei Chang.
\newblock Retrieval augmented code generation and summarization.
\newblock {\em arXiv preprint arXiv:2108.11601}, 2021.

\bibitem{gpt2}
Alec Radford, Jeff Wu, Rewon Child, David Luan, Dario Amodei, and Ilya
  Sutskever.
\newblock Language models are unsupervised multitask learners.
\newblock 2019.

\bibitem{t5}
Colin Raffel, Noam Shazeer, Adam Roberts, Katherine Lee, Sharan Narang, Michael
  Matena, Yanqi Zhou, Wei Li, and Peter~J Liu.
\newblock Exploring the limits of transfer learning with a unified text-to-text
  transformer.
\newblock {\em Journal of machine learning research}, 21(140):1--67, 2020.

\bibitem{reid2024gemini}
Machel Reid, Nikolay Savinov, Denis Teplyashin, Dmitry Lepikhin, Timothy
  Lillicrap, Jean-baptiste Alayrac, Radu Soricut, Angeliki Lazaridou, Orhan
  Firat, Julian Schrittwieser, et~al.
\newblock Gemini 1.5: Unlocking multimodal understanding across millions of
  tokens of context.
\newblock {\em arXiv preprint arXiv:2403.05530}, 2024.

\bibitem{ren2020codebleu}
Shuo Ren, Daya Guo, Shuai Lu, Long Zhou, Shujie Liu, Duyu Tang, Neel
  Sundaresan, Ming Zhou, Ambrosio Blanco, and Shuai Ma.
\newblock Codebleu: a method for automatic evaluation of code synthesis.
\newblock {\em arXiv preprint arXiv:2009.10297}, 2020.

\bibitem{generation2}
X.~Ren, X.~Ye, D.~Zhao, Z.~Xing, and X.~Yang.
\newblock From misuse to mastery: Enhancing code generation with
  knowledge-driven ai chaining.
\newblock In {\em 2023 38th IEEE/ACM International Conference on Automated
  Software Engineering (ASE)}, pages 976--987, Los Alamitos, CA, USA, sep 2023.
  IEEE Computer Society.

\bibitem{codellama}
Baptiste Rozi{\`e}re, Jonas Gehring, Fabian Gloeckle, Sten Sootla, Itai Gat,
  Xiaoqing Tan, Yossi Adi, Jingyu Liu, Tal Remez, J{\'e}r{\'e}my Rapin, Artyom
  Kozhevnikov, I.~Evtimov, Joanna Bitton, Manish~P Bhatt, Cristian~Cant{\'o}n
  Ferrer, Aaron Grattafiori, Wenhan Xiong, Alexandre D'efossez, Jade Copet,
  Faisal Azhar, Hugo Touvron, Louis Martin, Nicolas Usunier, Thomas Scialom,
  and Gabriel Synnaeve.
\newblock Code llama: Open foundation models for code.
\newblock {\em ArXiv}, abs/2308.12950, 2023.

\bibitem{ppo}
John Schulman, Filip Wolski, Prafulla Dhariwal, Alec Radford, and Oleg Klimov.
\newblock Proximal policy optimization algorithms.
\newblock {\em arXiv preprint arXiv:1707.06347}, 2017.

\bibitem{shen2023pangu}
Bo~Shen, Jiaxin Zhang, Taihong Chen, Daoguang Zan, Bing Geng, An~Fu, Muhan
  Zeng, Ailun Yu, Jichuan Ji, Jingyang Zhao, et~al.
\newblock Pangu-coder2: Boosting large language models for code with ranking
  feedback.
\newblock {\em arXiv preprint arXiv:2307.14936}, 2023.

\bibitem{inrelevant}
Freda Shi, Xinyun Chen, Kanishka Misra, Nathan Scales, David Dohan, Ed~H. Chi,
  Nathanael Sch\"{a}rli, and Denny Zhou.
\newblock Large language models can be easily distracted by irrelevant context.
\newblock In Andreas Krause, Emma Brunskill, Kyunghyun Cho, Barbara Engelhardt,
  Sivan Sabato, and Jonathan Scarlett, editors, {\em Proceedings of the 40th
  International Conference on Machine Learning}, volume 202 of {\em Proceedings
  of Machine Learning Research}, pages 31210--31227. PMLR, 23--29 Jul 2023.

\bibitem{shi2023replug}
Weijia Shi, Sewon Min, Michihiro Yasunaga, Minjoon Seo, Rich James, Mike Lewis,
  Luke Zettlemoyer, and Wen-tau Yih.
\newblock Replug: Retrieval-augmented black-box language models.
\newblock {\em arXiv preprint arXiv:2301.12652}, 2023.

\bibitem{rag-kg2}
Karthik Soman, Peter~W Rose, John~H Morris, Rabia~E Akbas, Brett Smith, Braian
  Peetoom, Catalina Villouta-Reyes, Gabriel Cerono, Yongmei Shi, Angela
  Rizk-Jackson, et~al.
\newblock Biomedical knowledge graph-enhanced prompt generation for large
  language models.
\newblock {\em arXiv preprint arXiv:2311.17330}, 2023.

\bibitem{ppo1}
Feifan Song, Bowen Yu, Minghao Li, Haiyang Yu, Fei Huang, Yongbin Li, and
  Houfeng Wang.
\newblock Preference ranking optimization for human alignment.
\newblock In {\em Proceedings of the AAAI Conference on Artificial
  Intelligence}, volume~38, pages 18990--18998, 2024.

\bibitem{arks}
Hongjin Su, Shuyang Jiang, Yuhang Lai, Haoyuan Wu, Boao Shi, Che Liu, Qian Liu,
  and Tao Yu.
\newblock Arks: Active retrieval in knowledge soup for code generation.
\newblock {\em arXiv preprint arXiv:2402.12317}, 2024.

\bibitem{ase-gold}
Ze~Tang, Jidong Ge, Shangqing Liu, Tingwei Zhu, Tongtong Xu, Liguo Huang, and
  Bin Luo.
\newblock Domain adaptive code completion via language models and decoupled
  domain databases.
\newblock In {\em 2023 38th IEEE/ACM International Conference on Automated
  Software Engineering (ASE)}, pages 421--433. IEEE, 2023.

\bibitem{tian2023chatgpt}
Haoye Tian, Weiqi Lu, Tsz~On Li, Xunzhu Tang, Shing-Chi Cheung, Jacques Klein,
  and Tegawend{\'e}~F Bissyand{\'e}.
\newblock Is chatgpt the ultimate programming assistant--how far is it?
\newblock {\em arXiv preprint arXiv:2304.11938}, 2023.

\bibitem{llama}
Hugo Touvron, Louis Martin, Kevin Stone, Peter Albert, Amjad Almahairi, Yasmine
  Babaei, Nikolay Bashlykov, Soumya Batra, Prajjwal Bhargava, Shruti Bhosale,
  et~al.
\newblock Llama 2: Open foundation and fine-tuned chat models.
\newblock {\em arXiv preprint arXiv:2307.09288}, 2023.

\bibitem{rag3}
Boxin Wang, Wei Ping, Lawrence McAfee, Peng Xu, Bo~Li, Mohammad Shoeybi, and
  Bryan Catanzaro.
\newblock Instructretro: Instruction tuning post retrieval-augmented
  pretraining.
\newblock {\em arXiv preprint arXiv:2310.07713}, 2023.

\bibitem{wang2023shall}
Boxin Wang, Wei Ping, Peng Xu, Lawrence McAfee, Zihan Liu, Mohammad Shoeybi,
  Yi~Dong, Oleksii Kuchaiev, Bo~Li, Chaowei Xiao, et~al.
\newblock Shall we pretrain autoregressive language models with retrieval? a
  comprehensive study.
\newblock {\em arXiv preprint arXiv:2304.06762}, 2023.

\bibitem{codet5+}
Yue Wang, Hung Le, Akhilesh~Deepak Gotmare, Nghi~DQ Bui, Junnan Li, and
  Steven~CH Hoi.
\newblock Codet5+: Open code large language models for code understanding and
  generation.
\newblock {\em arXiv preprint arXiv:2305.07922}, 2023.

\bibitem{codet5}
Yue Wang, Weishi Wang, Shafiq Joty, and Steven~CH Hoi.
\newblock Codet5: Identifier-aware unified pre-trained encoder-decoder models
  for code understanding and generation.
\newblock {\em arXiv preprint arXiv:2109.00859}, 2021.

\bibitem{rag-kg1}
Yilin Wen, Zifeng Wang, and Jimeng Sun.
\newblock Mindmap: Knowledge graph prompting sparks graph of thoughts in large
  language models.
\newblock In {\em Proceedings of the 62nd Annual Meeting of the Association for
  Computational Linguistics}, 2024.

\bibitem{xu2023recomp}
Fangyuan Xu, Weijia Shi, and Eunsol Choi.
\newblock Recomp: Improving retrieval-augmented lms with compression and
  selective augmentation.
\newblock {\em arXiv preprint arXiv:2310.04408}, 2023.

\bibitem{xu2022polycoder}
Frank~F. Xu, Uri Alon, Graham Neubig, and Vincent~Josua Hellendoorn.
\newblock A systematic evaluation of large language models of code.
\newblock In {\em Deep Learning for Code Workshop}, 2022.

\bibitem{crag}
Shi-Qi Yan, Jia-Chen Gu, Yun Zhu, and Zhen-Hua Ling.
\newblock Corrective retrieval augmented generation.
\newblock {\em arXiv preprint arXiv:2401.15884}, 2024.

\bibitem{prca}
Haoyan Yang, Zhitao Li, Yong Zhang, Jianzong Wang, Ning Cheng, Ming Li, and
  Jing Xiao.
\newblock {PRCA}: Fitting black-box large language models for retrieval
  question answering via pluggable reward-driven contextual adapter.
\newblock In Houda Bouamor, Juan Pino, and Kalika Bali, editors, {\em
  Proceedings of the 2023 Conference on Empirical Methods in Natural Language
  Processing}, pages 5364--5375, Singapore, December 2023. Association for
  Computational Linguistics.

\bibitem{ppo2}
Zheng Yuan, Hongyi Yuan, Chuanqi Tan, Wei Wang, Songfang Huang, and Fei Huang.
\newblock Rrhf: Rank responses to align language models with human feedback
  without tears.
\newblock {\em arXiv preprint arXiv:2304.05302}, 2023.

\bibitem{code-api}
Daoguang Zan, Bei Chen, Zeqi Lin, Bei Guan, Yongji Wang, and Jian-Guang Lou.
\newblock When language model meets private library.
\newblock {\em arXiv preprint arXiv:2210.17236}, 2022.

\bibitem{repocoder}
Fengji Zhang, Bei Chen, Yue Zhang, Jacky Keung, Jin Liu, Daoguang Zan, Yi~Mao,
  Jian-Guang Lou, and Weizhu Chen.
\newblock Repocoder: Repository-level code completion through iterative
  retrieval and generation.
\newblock {\em arXiv preprint arXiv:2303.12570}, 2023.

\bibitem{zhang2024codeagent}
Kechi Zhang, Jia Li, Ge~Li, Xianjie Shi, and Zhi Jin.
\newblock Codeagent: Enhancing code generation with tool-integrated agent
  systems for real-world repo-level coding challenges.
\newblock {\em arXiv preprint arXiv:2401.07339}, 2024.

\bibitem{zhang2023toolcoder}
Kechi Zhang, Huangzhao Zhang, Ge~Li, Jia Li, Zhuo Li, and Zhi Jin.
\newblock Toolcoder: Teach code generation models to use api search tools.
\newblock {\em arXiv preprint arXiv:2305.04032}, 2023.

\bibitem{generaion3}
Shun Zhang, Zhenfang Chen, Yikang Shen, Mingyu Ding, Joshua~B. Tenenbaum, and
  Chuang Gan.
\newblock Planning with large language models for code generation.
\newblock In {\em The Eleventh International Conference on Learning
  Representations}, 2023.

\bibitem{generaion1}
Tianyi Zhang, Tao Yu, Tatsunori Hashimoto, Mike Lewis, Wen-tau Yih, Daniel
  Fried, and Sida Wang.
\newblock Coder reviewer reranking for code generation.
\newblock In {\em International Conference on Machine Learning}, pages
  41832--41846. PMLR, 2023.

\bibitem{knn-tranx}
Xiangyu Zhang, Yu~Zhou, Guang Yang, and Taolue Chen.
\newblock Syntax-aware retrieval augmented code generation.
\newblock In {\em The 2023 Conference on Empirical Methods in Natural Language
  Processing}, 2023.

\bibitem{zhao2024ltgc}
Qihao Zhao, Yalun Dai, Hao Li, Wei Hu, Fan Zhang, and Jun Liu.
\newblock Ltgc: Long-tail recognition via leveraging llms-driven generated
  content.
\newblock {\em arXiv preprint arXiv:2403.05854}, 2024.

\bibitem{codegeex}
Qinkai Zheng, Xiao Xia, Xu~Zou, Yuxiao Dong, Shan Wang, Yufei Xue, Zihan Wang,
  Lei Shen, Andi Wang, Yang Li, et~al.
\newblock Codegeex: A pre-trained model for code generation with multilingual
  evaluations on humaneval-x.
\newblock {\em arXiv preprint arXiv:2303.17568}, 2023.

\bibitem{docprompting}
Shuyan Zhou, Uri Alon, Frank~F Xu, Zhiruo Wang, Zhengbao Jiang, and Graham
  Neubig.
\newblock Docprompting: Generating code by retrieving the docs.
\newblock {\em arXiv preprint arXiv:2207.05987}, 2022.

\end{thebibliography}
\end{document}